\documentclass{aa}
\usepackage{txfonts}
\usepackage{graphicx}
\usepackage{natbib}
\bibpunct{(}{)}{;}{a}{}{,}   

\begin{document}
\title{Dwarf elliptical galaxies in Centaurus A group: stellar populations in
AM~1339-445 and AM~1343-452\thanks{Based 
on observations collected at the European Southern Observatory, Paranal,
Chile, within the Observing Programme 073.B-0131, and on observations made
with the NASA/ESA {\it Hubble Space Telescope}, obtained from the data
archive at the Space Telescope Science Institute.  STScI is operated by the
Association of Universities for Research in Astronomy, Inc., under NASA
Contract NAS 5-26555.}
\thanks{Tables 2 and 3 are only available in electronic form
at the CDS  via anonymous ftp to cdsarc.u-strasbg.fr (130.79.128.5)
or via http://cdsweb.u-strasbg.fr/cgi-bin/qcat?J/A+A/../..}} 

\author{M.~Rejkuba\inst{1}
	\and G.~S.~Da Costa\inst{2}
	\and H.~Jerjen\inst{2}
	\and M.~Zoccali\inst{3}
	\and B.~Binggeli\inst{4}
}

\offprints{M. Rejkuba}

\institute{European Southern Observatory, Karl-Schwarzschild-Strasse
           2, D-85748 Garching, Germany\\
           E-mail: mrejkuba@eso.org
	\and Research School of Astronomy and Astrophysics, 
	     Institute of Advanced Studies, Australian National University, 
	     Cotter Road, Weston Creek, ACT 2611, Australia\\
	   E-mail: [gdc, jerjen]@mso.anu.edu.au
	\and Department of Astronomy and Astrophysics, Pontificia Universidad
	Cat\'olica de Chile, Vicu\~na Mackenna 4860, Santiago 22, Chile\\
	   E-mail: mzoccali@astro.puc.cl
	\and Astronomy Department, University of Basel, Venusstrasse 7, Basel,
	Switzerland\\
	   E-mail: binggeli@astro.unibas.ch
	}

\date{23 September 2005 / 23 November 2005}
\titlerunning{Stellar populations in AM~1339-445 and AM~1343-452}

\abstract{We study the red giant populations of two dE galaxies, 
AM~1339-445 and AM~1343-452, with the aim of investigating the number 
and luminosity of any upper asymptotic giant branch (AGB) stars present. 
The galaxies are members of the 
Centaurus~A group (D $\approx$ 3.8 Mpc) and are classified as outlying
(R $\approx$ 350 kpc) satellites of Cen~A\@. 
The analysis is based on near-IR photometry for individual 
red giant stars, derived from images obtained with ISAAC 
on the VLT. The photometry, along 
with optical data 
derived from WFPC2 images retrieved from the HST science archive, enable 
us to investigate the stellar populations of the dEs in the 
vicinity of the red giant branch (RGB) tip.  In both systems we find stars 
above the RGB tip, which we interpret as intermediate-age upper-AGB stars.
The presence of such stars is indicative of extended
star formation in these
dEs similar to that seen in many, but not all, dEs in the Local Group.
For AM~1339-445, the brightest of the upper-AGB stars have M$_{bol}$ 
$\approx$ --4.5
while those in AM~1343-452 have M$_{bol}$ $\approx$ --4.8 mag.  These
luminosities suggest ages of approximately 6.5 $\pm$ 1 and 4 $\pm$ 1 Gyr 
as estimates for the epoch 
of the last episode of significant star formation in these systems.  
In both cases the number of upper-AGB stars suggests that
$\sim$15\% of 
the total stellar population is in the form of intermediate-age stars,
considerably less than is the case for outlying dE satellites of the 
Milky Way such as Fornax and Leo~I.
}

\maketitle

%
%

\section{Introduction}
\label{sect:Intro}

Dwarf elliptical and  
dwarf spheroidal galaxies (we will refer to both classes as dwarf
elliptical (dE) galaxies from now on) are often assumed to have 
simple star formation histories because, at the present epoch, they 
generally lack neutral hydrogen and show no current or very recent 
star formation.  These properties separate them, morphologically,
from the gas-rich, star-forming dwarf irregular (dIrr) galaxies.  However, 
detailed studies of 
resolved stellar populations in Local Group (LG) dEs have revealed a 
surprising diversity of star formation histories, with some LG
dEs containing stars as young as $\sim$1 Gyr, or perhaps even younger 
\citep[e.g.\ Fornax and Leo~I; 
see reviews by][]{dacosta97,mateo98,vandenbergh99,vandenbergh00,grebel00}.

The dwarf galaxies in the LG follow a morphology-density relation  
in that the dEs are preferentially close to the Milky Way (MW) or M31, the 
two most massive galaxies in the LG, while almost all of the 
dIrr galaxies are located in the outskirts of the LG
in more isolated regions.  Further, among the MW and M31 dE 
satellite galaxies, there is a tendency for the systems that lie at larger 
distances to have larger intermediate-age (i.e.\ age $\approx$ 1--10 Gyr)
populations \citep[e.g.][]{vandenbergh94}.  This has led to the conjecture 
that proximity to a luminous galaxy, and the type of that galaxy, can
play an important role in defining dwarf galaxy evolution.  In particular, 
externally driven processes such as ram-pressure stripping by the hot
gaseous halo of a massive galaxy, and/or tidal effects, may control the rate
at which gas is lost from a dwarf system, and thus its star formation 
history. This approach has support from numerical simulations
\citep[e.g.][]{mayer+01b}. \citet{grebel+03} also argue that external
gas removal mechanisms are required to generate low-luminosity dEs, though they
suggest that the transition-type dwarfs, i.e.\ the dwarfs classified as 
dIrr/dE, are the best model for low luminosity dEs that have not lost their
gas, rather than dIrrs.

However, the model in which the evolution of a dwarf system is halted by
externally induced gas loss, does not easily explain 
the existence of LG dEs like Tucana and Cetus.  These dwarfs show very little, 
if any, evidence for extended star formation despite having isolated 
locations far
from the MW or M31 \citep[e.g.][]{dacosta98,sarajedini+02}.  If these dEs
have always remained isolated, then an internal process must have been 
responsible for the apparent complete gas loss at early times.

In order to explore further the importance of external vs.\ internal 
processes in
driving the star formation in these small systems, it is necessary to study in
detail dE galaxies in environments different from that of the LG\@. 
The relatively nearby Centaurus~A Group (D $\approx$ 3.8 Mpc) is one such 
environment.  This group, which has the unusual giant E~galaxy Cen~A as its
single dominant member, is more compact than the LG and it probably contains 
perhaps twice as many galaxies.  For instance, 
\citet{k+02} list 13 galaxies that are most likely within a 600 kpc radius 
of Cen~A, and which are brighter than M$_{B}$ $\approx$ --12.
Besides Cen~A itself, these include the giant galaxies NGC~4945 and NGC~5102 
as well as 10 dwarf
galaxies, 6 of which are early-type and 4 late-type.  This catalogue is
likely to be significantly incomplete.
In contrast, in the Local Group, for an approximately equivalent volume 
centered on the LG barycentre and with an equivalent magnitude cutoff, the
compilation of \citet{vandenbergh00} yields a (complete) sample of 13 galaxies:
4 large galaxies (MW, M31, M33 and the LMC), 6 early-type dwarfs (M32, 
NGC~205, NGC~185, NGC~147, Fornax and Sagittarius) and 3 late-type dwarfs
(SMC, IC10 and IC1613).  Thus it is 
likely that the Cen~A group has provided a significantly different environment
for its dwarf members than has the LG\@.

We have begun a program to study the red giant populations of 
the dE galaxies in the Cen~A group, with the ultimate aim of investigating the
extent to which star formation history indicators correlate with distance 
from the dominant galaxy of the group, Cen~A\@.  Specifically, we will
investigate the number and luminosity of upper-AGB stars in the dE galaxies.
Upper-AGB stars are stars
with sufficient mass to evolve to luminosities above the RGB tip, and their
presence, provided the system is relatively metal-poor 
($\langle$[Fe/H]$\rangle$
$\leq$ --1.0), as is the case for all but the most luminous dEs, is an 
unambiguous indicator of the existence of an intermediate-age 
($\sim$1--10 Gyr) population.  The luminosity of the brightest upper-AGB 
stars is also a measure of the age of the youngest intermediate population of
significance.  Because of their cool
effective temperatures, these upper-AGB stars are best studied in the 
near-infrared.
Indeed it was near-IR observations that provided the first evidence for the 
diversity of star formation histories among the MW dSph companions 
\citep[cf.][]{aaronson+mould80}.

Using the ISAAC near-IR array at the ESO Very Large Telescope (VLT) we have
obtained $J_s$ and $K_s$-band images of 14 Cen~A 
group dwarf galaxies. The full data set, together with description of
the reduction and analysis techniques, 
will be presented in a forthcoming paper (Rejkuba et al.,  
in preparation).
We present here the first results of our program -- analysis of the resolved 
red giant stellar populations of two Cen~A group dE galaxies, AM~1339-445 
(KK 211) and AM~1343-452 (KK 217). 
For these two galaxies it is possible to supplement the near-IR data with 
$V$ and $I$ band WFPC2 images from the HST science archive, permitting 
estimates of the distances and of the average metallicity of their stars.

In Table~\ref{tab:target_character} we summarize the fundamental 
parameters of the two dwarf ellipticals. The last column gives 
the literature reference for the tabulated data.

Using the integrated magnitudes of 
\citet{JBF00} and the reddenings and moduli adopted here (see
Table~\ref{tab:target_character} and following
sections) the absolute blue magnitudes of these two galaxies are 
M$_{B}=-11.9$ and $-10.8$, respectively, with colours $(B-R)_0=1.38$ and 
$1.35$, typical for dE galaxies \citep[e.g.][]{evans+90}. 
Neither galaxy is detected in the HIPASS survey: the 3-$\sigma$ upper
limits on their H{\small I} contents correspond to 
$2.8 \times 10^6 M_\odot$  and 
$3.1 \times 10^6 M_\odot$, respectively, for an assumed 10 kms$^{-1}$ line 
width and detection limits of the survey from \citet{barnes+01}. 
\citet{k+02}
list line-of-sight distances for the two galaxies based on the $I$ magnitude
of the red giant branch tip (see also Sect.\ \ref{sect:results_optical}).
With those distances, the angular separations from Cen~A, and a distance
for Cen~A of 3.84 Mpc \citep{rejkuba04}, the true separation of AM~1339-445 
from Cen~A is $\sim$390 kpc, while that of AM~1343-452 is approximately 
320 kpc. \citet{k+02} label both dwarfs as companions of Cen~A\@.  
The distances of both these 
dEs from Cen~A are therefore somewhat larger than those of the outer dE
satellites Leo~I and Leo~II from the MW, and those of the outer
dE satellites And~II, And~VI (Peg) and And~VII (Cas) from M31.  

\begin{table}
\caption{Fundamental parameters of the two target dE galaxies in Cen A group.}
\label{tab:target_character}
\centering
\begin{tabular}{lccl}
\hline
\hline
Name               & AM~1339-445  & AM~1343-452 &\\
\hline
$\alpha_{J2000.0}$ & 13:42:05.8   &  13:46:18.8 &\\
$\delta_{J2000.0}$ & $-$45:12:21  & $-$45:41:05 &\\
Type               & dE           & dE          & 1\\
$B_T$ (mag)        & $16.32$        & $17.57$       & 1\\
$R_T$ (mag)        & $14.76$        & $16.02$       & 1\\
$(B-R)_0^T$ (mag)  & $1.38$           & $1.35$   & 1 \\
$r_{\mathrm{eff},R}$ (arcsec)& $23.8$   & $14.7$    & 1\\
$\langle \mu \rangle_{\mathrm{eff},R}$& $23.63$  & $23.85$ & 1\\
E($B-V$) (mag)     & $0.111$        & $0.121$    & 2 \\
$(m-M)_0$ (mag)   & $27.87 \pm 0.27$ & $27.99 \pm 0.37$ & 1 (SBF)\\
   & $27.77 \pm 0.21$ & $27.92 \pm 0.25$ & 3 (TRGB)\\
   & $27.74 \pm 0.20$ & $27.86 \pm 0.20$ & 4 (TRGB)\\
$\langle \mathrm{[Fe/H]} \rangle $ &$-1.4 \pm 0.2 $ & $-1.6 \pm 0.2$& 4\\
\hline
References: &\multicolumn{3}{l}{1: \citet{JFB00}; 2: \citet{schlegel+98};}\\
&\multicolumn{3}{l}{3: \citet{k+02}; 4: this work}
\end{tabular}
\end{table}

The paper is organised as follows.  In the following section the
observations and reductions are described, first for the near-IR data
and then for the WFPC2 data.  The third section outlines the analysis of
the resulting colour-magnitude diagrams (CMDs) for the two dwarfs.  The
fourth section discusses the results, which are summarised in the final 
section.   
A preliminary description of this work has appeared in \citet{dacosta05}
and \citet{rejkuba+05}.

%
%

\section{Observations and reductions}
\label{sect:O+R}

The observations and the subsequent photometry of the ISAAC frames 
are summarized in
Sect.~\ref{sect:ISAAC-data}.  More details of
the reduction and analysis techniques will be presented in a forthcoming 
paper (Rejkuba et al., in preparation), together with the photometry of the 
remaining 12 Cen~A group dwarf galaxies in our sample.

The optical images of our program galaxies come from the HST WFPC2 `snapshot' 
programme 8192 (PI: Seitzer), and have been discussed in \citet{k+02}. These 
authors were primarily interested in distances,
derived from the tip of the RGB, to establish, or confirm,
membership in the Cen~A group for a sizeable sample of dwarfs of all types.  
For both dEs discussed here, however, \citet{JFB00} had
previously established Cen~A group membership via a distance determined
with the surface brightness fluctuation method.  Since \citet{k+02}
did not publish their photometry for individual stars, we retrieved the
WFPC2 data frames from
the HST science archive and carried out our own photometric analysis.
This is described in Sect.~\ref{sect:HST-data}. We then repeat the 
distance analysis and use the CMDs to infer the average metallicity of 
the dE stars and to search for candidate upper-AGB stars. 

The combination of the WFPC2 and ISAAC photometry is described in  
Sect.~\ref{sect:ISAAC-HST-combine} and the final catalogue is
available as an electronic table from CDS\@.
We note that in order to correct our photometry for Galactic extinction, we 
use the following 
$\mathrm{E(B-V)}$ values throughout the paper  (cf.\
Table~\ref{tab:target_character}):
$\mathrm{E(B-V)_{AM1339-445}}=0.111$ and $\mathrm{E(B-V)_{AM1343-452}}=0.121$
\citep{schlegel+98}. We adopt 
A$_{V}$ = 3.3E$(B-V)$ and the E$(V-I)$, E$(B-V)$ relation from \citet{dean+78},
while the \citet{cardelli+89} extinction law is used to calculate 
A$_{J}$ and A$_{K}$ values.

\subsection{ISAAC data}
\label{sect:ISAAC-data}

The Near-IR observations of AM 1339-445 and AM 1343-452 
were taken in service mode 
using the short wavelength arm of the ISAAC instrument on the Antu (UT1) 
Very Large Telescope (VLT) at ESO Paranal Observatory. The field-of-view is 
$2\farcm 5 \times 2\farcm 5$ and the pixel scale $0\farcs 148$.

Each galaxy was observed once in the $J_s$ and twice in the $K_s$-band. 
AM 1339-445 was observed on July 19, 2004 in the $J_s$ band, and the two 
$K_s$-band epochs were secured on May 10 and May 14, 2004. The $J_s$ 
and one $K_s$ epoch of AM 1343-452 were taken on June 27, 2004, and the other
$K_s$ observation on May 14, 2004. Each observation consisted of a sequence of
jittered short exposures. The total exposure times were 2352 sec 
for each $K_s$ observation and 2100 sec for $J_s$.

The standard procedure in reducing infrared (IR) data consists of dark
subtraction, flat-field correction, sky subtraction, 
registering and combining the images. 
The details of the adopted data reduction procedures within IRAF are 
described in  
\citet{rejkuba+01}.
At the end of the reduction all the images taken within each individual 
jittered sequence were combined. From now on, when
we refer to an image in the $J_s$, or $K_s$-band, we always mean these 
combined sequences. All the images have very high resolution thanks to 
excellent seeing. For example,  on the $K_s$ band images of both galaxies taken on 
May 14, 2004, we measure FWHM of only $0\farcs 33$, while the $J_s$-band 
images have slightly worse resolution, with seeing of $0\farcs58$.

Point-spread-function (PSF) fitting photometry on the reduced images was done 
using the suite of DAOPHOT, ALLSTAR 
and ALLFRAME programmes \citep{stetson87,stetson94}.
We combined the $J_s$ and $K_s$ images of each galaxy to derive the master
star list, which was then used to obtain photometry from the $J_s$ and the two 
$K_s$ images. 
The final photometric catalogue for each galaxy contains all the sources that 
could be measured in both $J_s$ and at least one $K_s$-band image and that had 
photometric errors, measured by ALLFRAME, smaller than 0.3 mag, and PSF fitting
quality parameters $\chi \leq 1.5$ and $|\mathrm{sharpness}| < 2$. 
After this selection
the catalogues contain 826 stars in the field centered on AM~1339-445 and 743
stars in AM~1343-452. 

\begin{figure}
\centering
\resizebox{\hsize}{!}{
\includegraphics[angle=270]{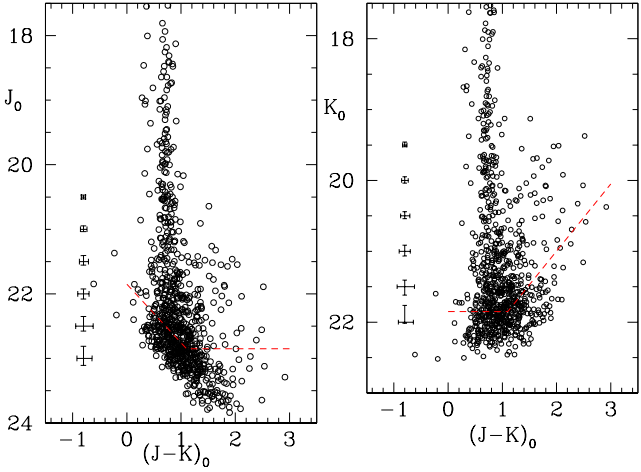}
}
\caption[]{AM 1339-445 near-IR CMDs. Dashed lines indicate 50\% completeness limits, 
and the typical error-bars are indicated on the left side. For example, at 
$K_0=21.5$ and $(J-K)_0=1.0$ they amount to $\sigma_K=0.2$ 
and $\sigma_{(J-K)}=0.3$, while one magnitude brighter, at $K_0=20.5$, and
at the same colour, they are $\sigma_K=0.09$ and $\sigma_{(J-K)}=0.15$.
}
	
   \label{fig:AM1339JKcmds}
\end{figure}

\begin{figure}
\centering
\resizebox{\hsize}{!}{
\includegraphics[angle=270]{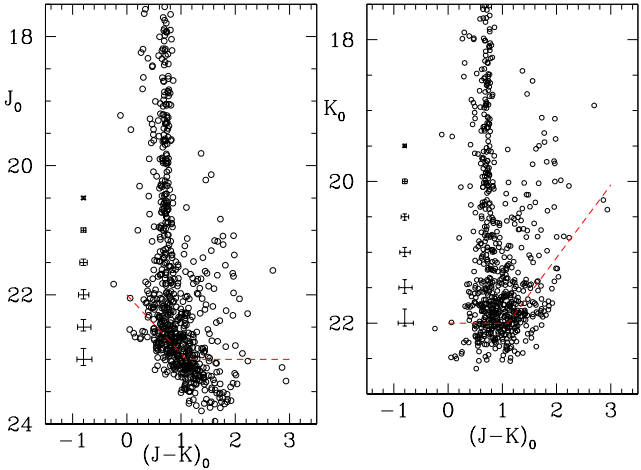}
}
\caption[]{AM 1343-452 near-IR CMDs. Dashed lines indicate 50\% completeness limits,
 and the typical error-bars are indicated on the left side. For example, at 
$K_0=21.5$ and $(J-K)_0=1.0$ they amount to $\sigma_K=0.15$ 
and $\sigma_{(J-K)}=0.25$, while one magnitude brighter, at $K_0=20.5$, and
at the same colour, they are $\sigma_K=0.08$ and $\sigma_{(J-K)}=0.12$. 
	}
   \label{fig:AM1343JKcmds}
\end{figure}

Photometric calibration was done by applying the zero point difference between 
our instrumental magnitude measurements and magnitudes of stars found in common 
with the 2MASS  all-sky catalog of point sources \citep{cutri+03}, as
previous experience with this instrumental system has shown that any colour
terms are negligible. 
The respective $J_{0}$ vs.\ $(J-K)_{0}$ and $K_{0}$ vs.\ $(J-K)_{0}$ CMDs
are shown in Fig.~\ref{fig:AM1339JKcmds} and \ref{fig:AM1343JKcmds}.
The brighter parts of the CMDs are dominated by field stars, many of 
which have similar $(J-K)_{0}$ colours. Stars found in common with the 2MASS catalog,
and used for calibration, are among the brightest of these field stars. 
The contribution to the CMDs
from the galaxy stars are found at the fainter magnitudes.

Completeness and magnitude errors were measured using  
artificial-star tests, which are discussed in more detail in 
the paper that describes the whole dataset (Rejkuba et al., in preparation).
Specifically, the tests show that for these galaxies, the completeness and the 
magnitude errors as a function of magnitude vary very little across the 
field, indicating that there are no problems with crowding in the centers 
of the images. The 50\% 
completeness limit is reached at $J_0 = 22.8$ and $K_0=21.8$ in 
AM~1339-445, and $J_0 = 23.0$ and $K_0=22.0$ in AM~1343-452.
These completeness limits are indicated on CMDs in Fig.~\ref{fig:AM1339JKcmds} 
and \ref{fig:AM1343JKcmds} with dashed lines. Typical error-bars, computed as 
average difference between input and recovered magnitudes as a function of 
input magnitude in completeness simulations, are also plotted. 
The errors for colour typical of RGB stars and magnitudes close 
to the RGB tip as well as one 
magnitude brighter are quoted in the captions of these figures.

\subsection{HST data}
\label{sect:HST-data}

The HST WFPC2 data for each galaxy consists of single 600 sec exposures
through the $F606W$ (`Wide-$V$') and $F814W$ (`Wide-$I$') filters.  The
pipeline-processed image files and the accompanying data quality files
were retrieved from the HST science archive.  In both cases the galaxy
was centered on the WF3 chip.  The images were analysed with the 
December 2004 version (version 1.1.7a) of Dolphin's HSTphot package
\citep{HSTphot}.  For each galaxy,
the data quality files were first used to mask bad pixels and other 
defects in the images.  The `sky' image was then computed with the
{\it getsky} routine and the {\it crmask} code used in conjunction with
both images to remove cosmic rays.  Finally, the {\it HSTphot} code was 
run to produce $V$ and $I$ photometry for all objects detected on
both images.  This code includes updated CTE corrections and zeropoint
calibrations \citep[cf.][]{dolphin00}.  Because of the small number of 
stars on the PC1 frames, the data from that chip were omitted from the 
further analysis.

\begin{figure}
\centering
\resizebox{\hsize}{!}{
\includegraphics[angle=0]{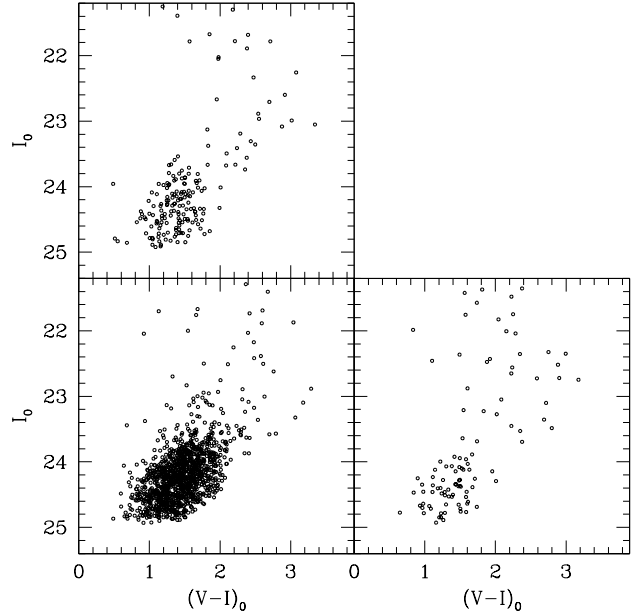}
}
\caption[]{Reddening corrected CMDs for AM 1339-445 from the WFPC2 data. 
WF2 is the upper-left, WF3 is the lower-left and WF4 is the lower-right.
	}
   \label{fig:HSTcmds1339}
\end{figure}

\begin{figure}
\centering
\resizebox{\hsize}{!}{
\includegraphics[angle=0]{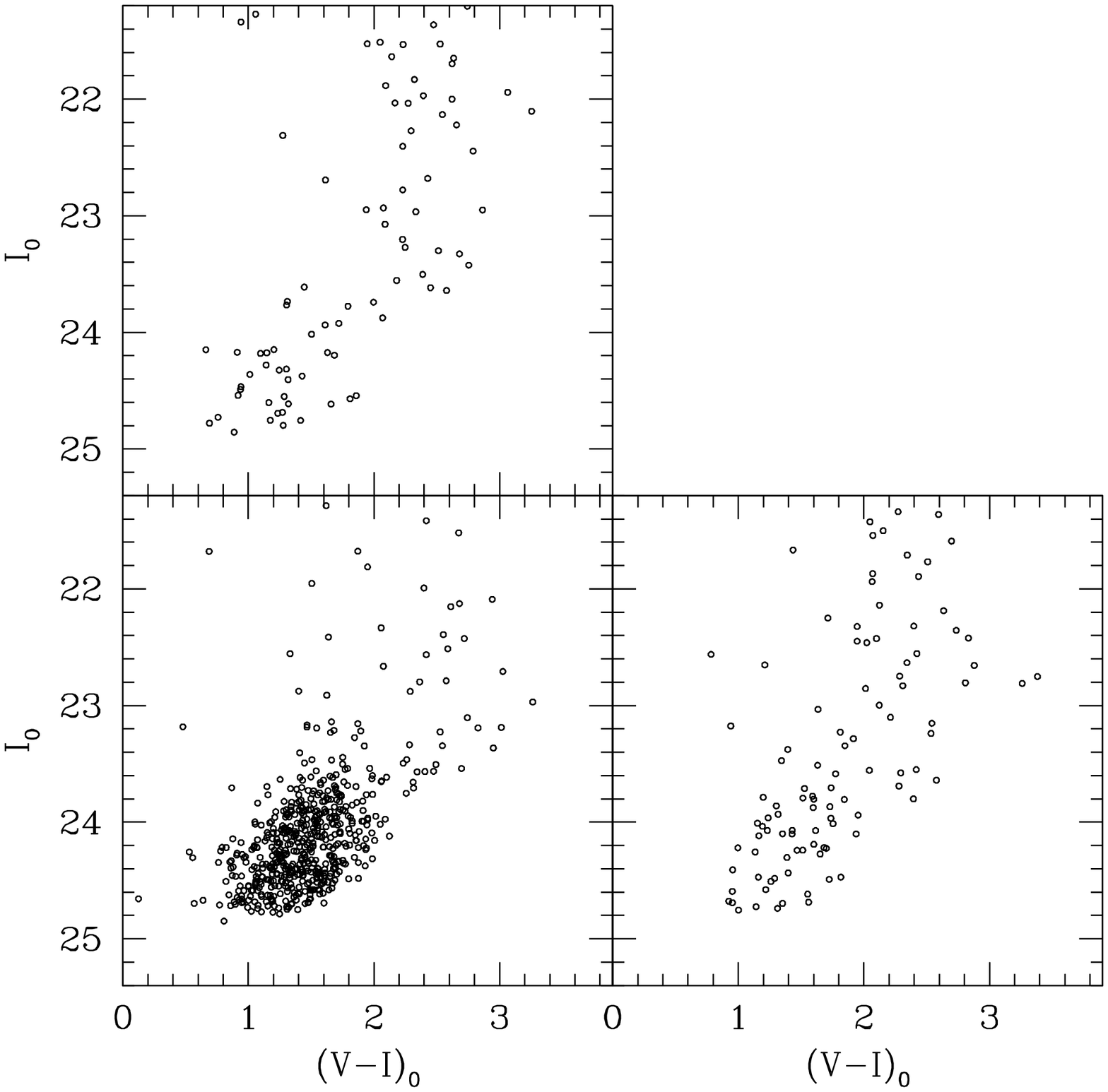}
}
\caption[]{Reddening corrected CMDs for AM 1343-452 from the WFPC2 data. 
WF2 is the upper-left, WF3 is the lower-left and WF4 is the lower-right. 
	}
   \label{fig:HSTcmds1343}
\end{figure}

The output of {\it HSTphot} contains a number of parameters for each
detected object.  These include a classification, $\chi$, sharpness and
roundness parameters, errors in the final magnitude, and estimates of the
effect of crowding.  In order to produce as
clean a CMD as possible, only objects with a classification of 1 (`good
star') were retained.  Further, stars with discrepant $\chi$, sharpness,
roundness and errors for their magnitude were discarded.  Similarly, any
star with a crowding parameter exceeding 0.05 mag was excluded. The
reddening corrected CMDs for the two galaxies are shown in 
Fig.~\ref{fig:HSTcmds1339}
for AM~1339-445, and Fig.~\ref{fig:HSTcmds1343} for AM~1343-452.  
It is immediately
apparent that the galaxy stars are primarily confined to the WF3 chip,
though there are indications that galaxy stars also populate the other 
chips to some extent, particularly WF2 for AM~1339-445.

\subsection{Combined data}
\label{sect:ISAAC-HST-combine}

\begin{figure}
\centering
\resizebox{\hsize}{!}{
\includegraphics[angle=0]{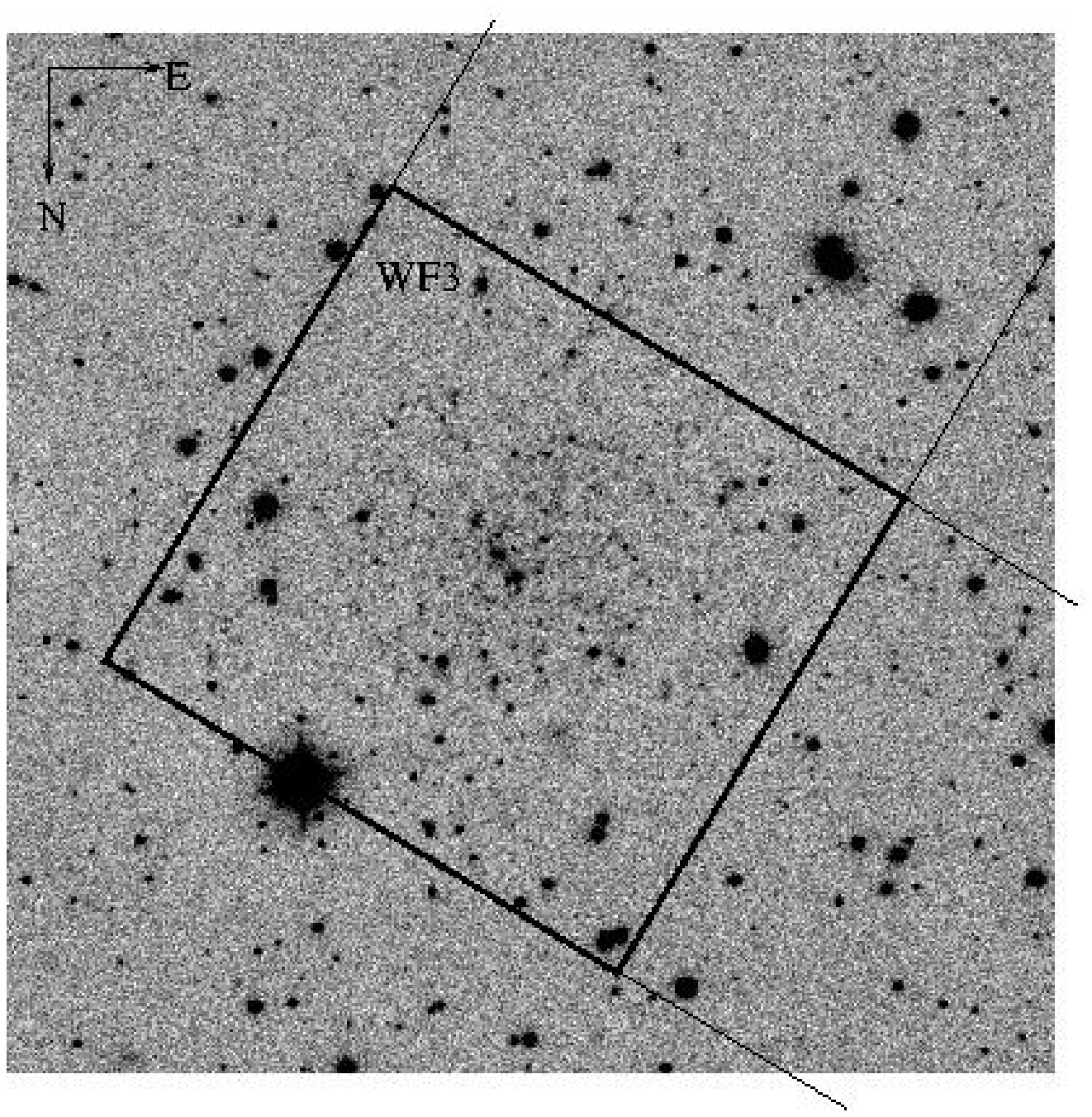}
}
\caption[]{The ISAAC K-band image of AM~1339-445 taken on May 10, 2004, 
with the WF3 field of view indicated with thick lines. The thinner 
lines indicate the overlap with the WF2 and WF4 chips. Seeing measured 
on this ISAAC image is $0\farcs50$.
	}
   \label{fig:AM1339_ISAAC_HST}
\end{figure}

\begin{figure}
\centering
\resizebox{\hsize}{!}{
\includegraphics[angle=0]{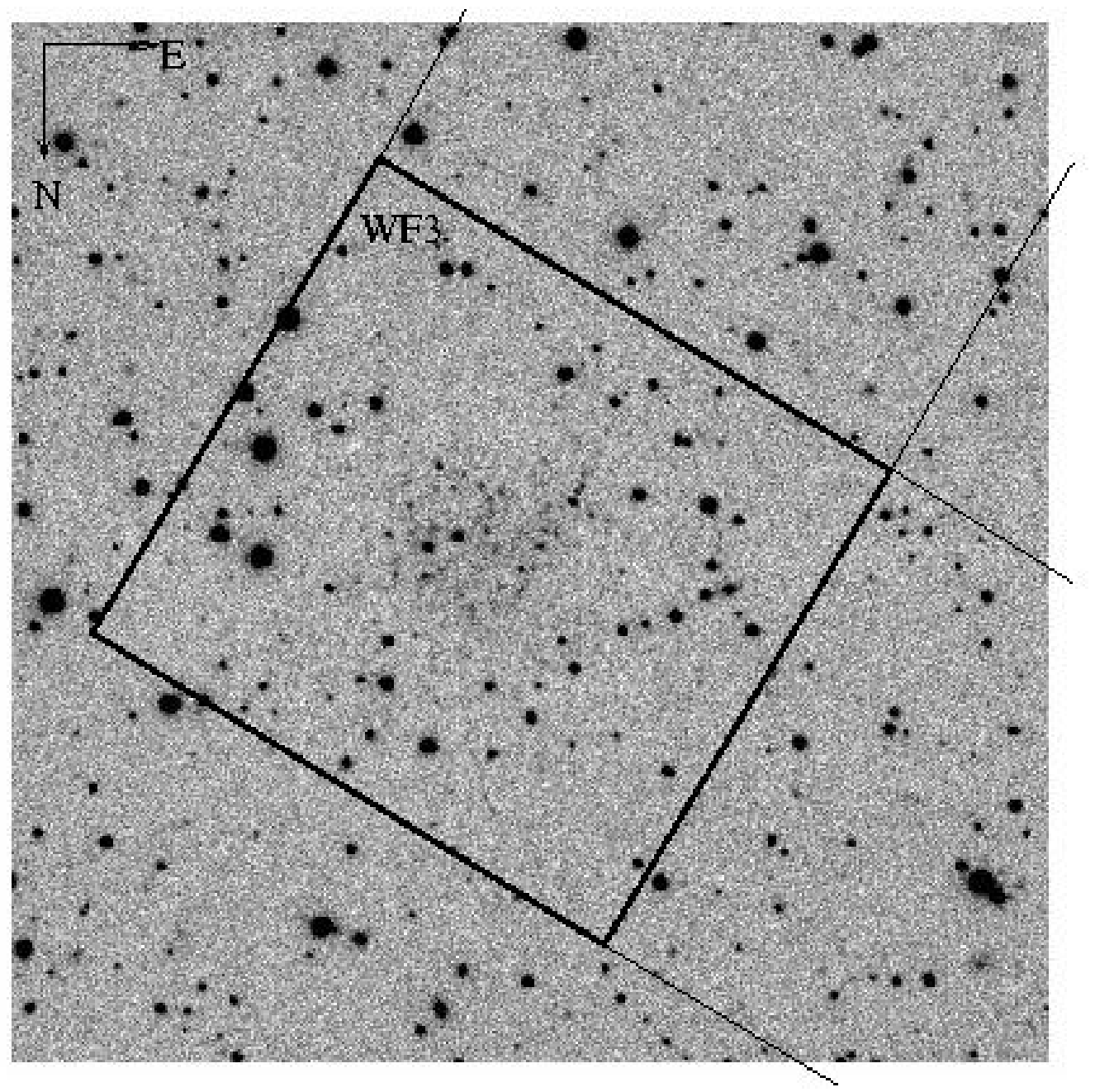}
}
\caption[]{The ISAAC K-band image of AM~1343-452 taken on June 27, 2004, 
with the WF3 field of view indicated with thick lines. The thinner 
lines indicate the overlap with the WF2 and WF4 chips. Seeing measured 
on this ISAAC image is $0\farcs58$.
	}
   \label{fig:AM1343_ISAAC_HST}
\end{figure}

\begin{table*}
\caption{Combined WFPC2 and ISAAC measurements in $V$, $I$, $J_s$ and the
average $K_s$ magnitude of all the stars in AM~1339-445  without
extinction correction. For each magnitude in parenthesis we give magnitude errors
as given by photometric package, HSTphot and DAOPHOT. Here we give only the first three entries,
and the complete table is available through the CDS.}
\label{tab:allmchstars1339}
\centering
\begin{tabular}{lrrcccc}
\hline
\hline
ID         & X(WF3)  & Y(WF3)& $V(\pm \sigma)$&$I(\pm \sigma)$&$J_s(\pm \sigma)$ &$K_s(\pm \sigma)$ \\
\hline
226 &151.61 & 731.13& 21.486 (0.010) & 19.551 (0.007) &  18.29 (0.01)&  17.49 (0.01) \\
234 &129.33 & 701.22& 26.027 (0.126) & 24.248 (0.121) &  22.96 (0.11)&  22.09 (0.26) \\
248 &140.59 & 689.67& 21.340 (0.007) & 19.541 (0.007) &  18.36 (0.01)&  17.59 (0.01) \\
... &    &    &            &            &           &           \\
\hline		
\end{tabular}	   									    
\end{table*}				   									    

\begin{table*}
\caption{Combined WFPC2 and ISAAC measurements in $V$, $I$, $J_s$ and the
average $K_s$ magnitude of all the stars in AM~1343-452 without
extinction correction. For each magnitude in parenthesis we give magnitude errors
as given by photometric package, HSTphot and DAOPHOT. Here we give only the first three entries,
and the complete table is available through the CDS.}
\label{tab:allmchstars1343}
\centering
\begin{tabular}{lrrcccc}
\hline
\hline
ID         & X(WF3)  & Y(WF3)& $V(\pm \sigma)$&$I(\pm \sigma)$&$J_s(\pm \sigma)$ &$K_s(\pm \sigma)$ \\
\hline
825 &363.97 & 755.06& 26.114 (0.157) & 24.365 (0.133) &  23.06 (0.14) & 21.86 (0.13) \\
877 &425.34 & 758.28& 23.315 (0.023) & 20.553 (0.012) &  18.73 (0.01) & 17.90 (0.01) \\
890 &219.26 & 627.35& 26.636 (0.194) & 23.776 (0.081) &  22.49 (0.10) & 21.36 (0.09) \\
... &    &    &             &             &           &            \\
\hline		
\end{tabular}	   									    
\end{table*}

While the total WFPC2 field is comparable to that of the ISAAC data,
it is evident from Figs.~\ref{fig:HSTcmds1339} and \ref{fig:HSTcmds1343} 
that the bulk of the red giants are contained on the WF3 chip. 
The centres of
the two galaxies are in the centre of the ISAAC and of the WF3 chip, while
the overlap between WF2 and WF4 chips and the ISAAC field of
view is much smaller, as can be seen from Figs.~\ref{fig:AM1339_ISAAC_HST} 
and \ref{fig:AM1343_ISAAC_HST}.
Consequently, we have decided to match only the WF3 $VI$ photometry 
with the ISAAC data. 

The initial coordinate transformations between the WF3 and ISAAC images were 
calculated using the IRAF tasks
{\it geomap} and {\it geoxytran}. The transformations were then refined and the
combined catalogues made using the DAOMASTER programme \citep{stetson87}.
The final optical-near-IR photometry lists contain 277 and 190 stars 
in AM~1339-445 and AM~1343-452, respectively. The complete tables of all 
the matched sources are published in the electronic version of the paper 
and are available through CDS\@.  We show in 
Tables~\ref{tab:allmchstars1339} and \ref{tab:allmchstars1343} 
the photometry for the first three stars in both galaxies. 

%
%
\section{Analysis}

\subsection{Foreground contamination}
\label{sect:contamination}
\begin{figure}
\centering
\resizebox{\hsize}{!}{
\includegraphics[angle=270]{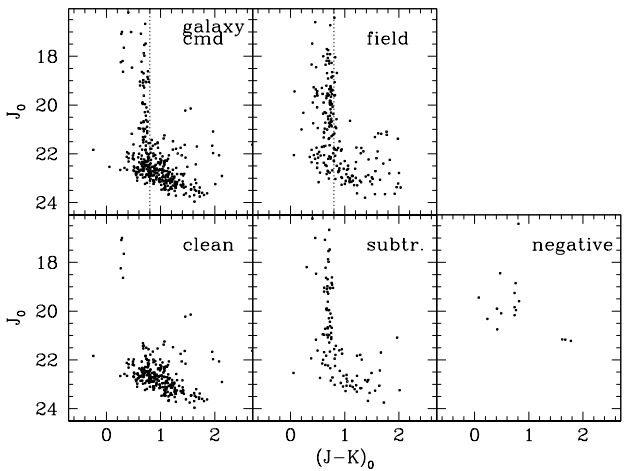}
}  
\caption[]
	{Decontamination of the CMD for AM~1343-452. In the upper panels are 
the observed CMDs for the galaxy (left) and field (center) areas. In the
bottom panels from left to right there are: the cleaned galaxy CMD (clean), 
the CMD made of stars subtracted statistically from the galaxy area that had 
colours and magnitudes similar to those from the field area (subtracted), 
and the CMD for stars from the field 
area that had no counterpart in the galaxy area (negative).

	}
   \label{fig:decontamination}
\end{figure}

As is evident in Figs.~\ref{fig:AM1339JKcmds} and \ref{fig:AM1343JKcmds}, 
the relatively low galactic latitude of the dE fields results in 
a relatively large number of foreground Galactic stars contaminating the
CMDs.  The Milky Way's old disk turn-off stars populate the vertical 
sequence around $(J-K)_0 \simeq 0.35$, fainter red giant branch and red clump 
stars are found along the most populated sequence at $(J-K)_0 \simeq 0.65$, 
and the low mass dwarfs with 
$\mathrm{M} \la 0.6~M_\odot$ have $(J-K)_0 \simeq 0.9$ \citep{marigo+03}. 
It is possible also that there are a few contaminating background 
compact galaxies.  
However, almost all the foreground stars are have colours 
bluer than $(J-K)_0 < 0.9$, and in the near-IR CMDs they are
well separated from most of the cool giants belonging to the galaxy. 

In order to decontaminate the near-IR CMDs we proceeded as follows. 
First, we determine the galaxy and the field areas on the ISAAC array. The 
former corresponds to the area where most of
the stars would belong to the target galaxy, while the latter contains mostly
foreground stars or compact background objects. The stellar radial density
distribution was obtained by counting stars in 
concentric rings centered on the galaxy. The galaxy region was assumed 
to extend inwards from the
level where the stellar density reached 20\% of the maximum, while for the
field region, we selected the area outwards from the point where the
density is 10\% of the central value. 
While it is possible that some galaxy stars are present in the 
field area, their contribution should be small.

\begin{figure}
\centering
\resizebox{\hsize}{!}{
\includegraphics[angle=0]{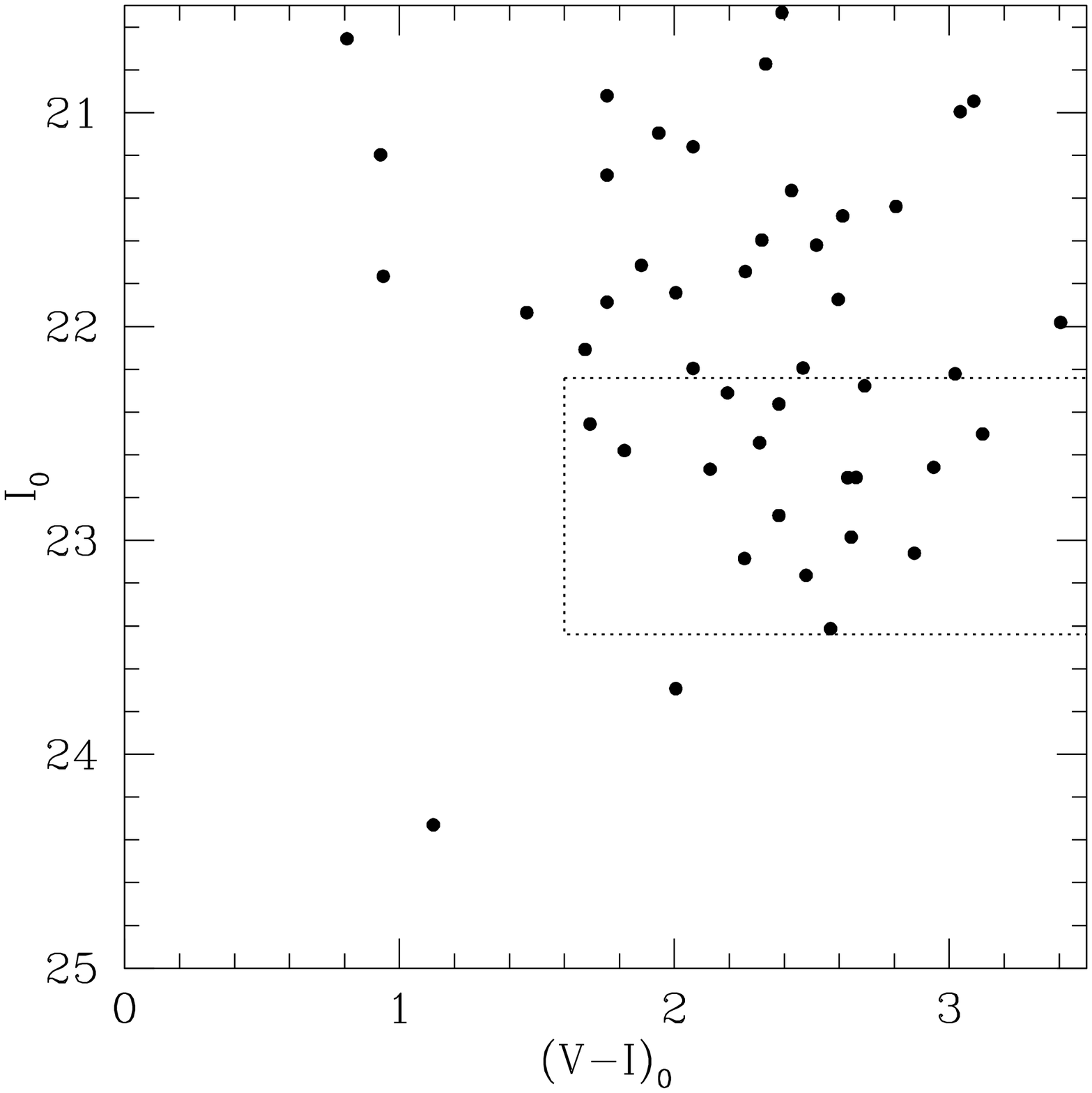}
\includegraphics[angle=0]{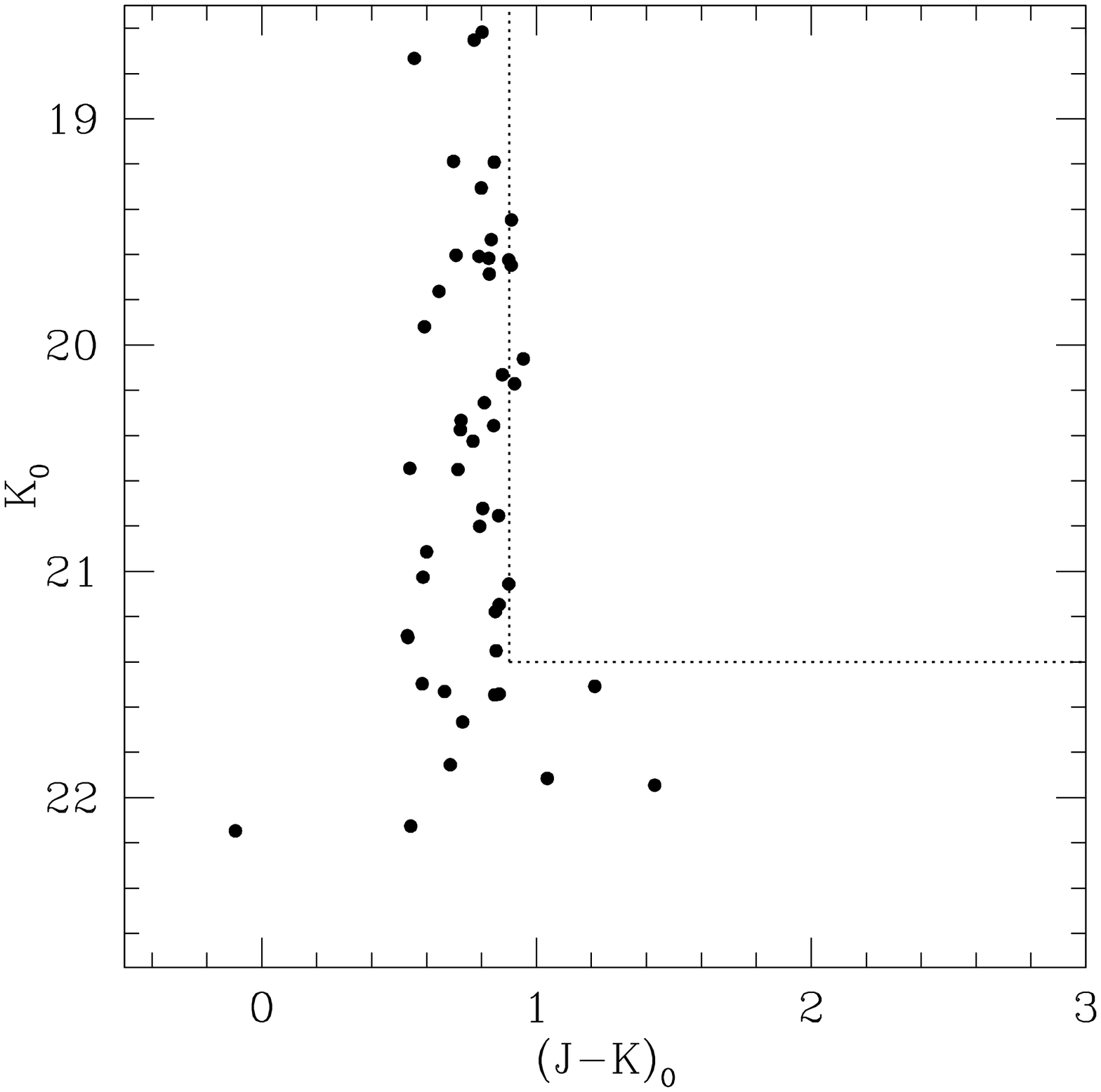}
}  
\caption[]
	{
	Example of one simulation of the expected Milky Way foreground stars in
the field of view of our combined dataset for AM~1339-445 from the Galaxy model 
of the Besan\c{c}on group \citep{robin+03}. The magnitudes of simulated stars
include realistic errors and correction for incompleteness. 
The dotted lines indicate the boxes used to search for upper-AGB star candidates.}
   \label{fig:besancon}
\end{figure}

Taking into account the relative sizes of galaxy and field areas, we then
subtracted, for each star found in the field area, a star with similar colour 
and magnitude in the galaxy area. The maximum allowed difference in colour 
between the two stars takes into account the photometric errors, while the 
maximum allowed difference in magnitude could be larger to allow for
statistical fluctuations in the luminosity distribution of the field stars.
The subtraction of a star in the galaxy area depends on whether there is a star
with similar colour and magnitude in the field area, and the probability 
of its subtraction
is proportional to the ratio of galaxy and field areas. We cannot
know which stars really belong to galaxy and which to field and thus this
decontamination process is necessarily not unique.  However, the main features
of the subtracted CMDs are robust with respect to different runs of the
subtraction code.
An example of the decontamination process is shown in 
Fig.~\ref{fig:decontamination} for AM~1343-452. In the upper panels we plot 
the observed CMDs for the galaxy (left) and field (center) areas. In the
bottom panels we plot from left to right: the cleaned galaxy CMD (clean), 
the CMD made of stars subtracted statistically from the galaxy area that had 
colours and magnitudes similar to those from the field area (subtracted), 
and finally on the right the colours and magnitudes of stars from the field 
area that had no counterpart in the galaxy area (negative). The 
occurrence of the ``negative stars'' is not unexpected given likely 
statistical fluctuations in the colour and magnitude distribution of the 
stars in the field region. 

In the optical, thanks to the orientation of the HST field of view (see
e.g.~Fig.~\ref{fig:AM1339_ISAAC_HST} and \ref{fig:AM1343_ISAAC_HST}), 
larger areas at larger distances from the galaxy centres could be used to 
estimate the field contamination.  For AM~1339-445 the field area is generated 
from the stars that lie in the halves of the WF2 and WF4 chips furthest from 
WF3, while for AM~1343-452 it is generated by averaging the numbers of stars 
on the WF2 and WF4 chips.

Unfortunately, because of the different fields-of-view, a similar procedure 
could not be employed to decontaminate the combined optical-IR diagrams. 
However, simulations of the Galactic foreground, using the Besan\c{c}on 
group model \citep{robin+03},  can be used to estimate the possible
contamination of the region of the CMDs where upper-AGB candidates 
are expected to be found. An example of a simulation is 
shown in Fig.~\ref{fig:besancon}, where we note that we have added the
simulated stars into our images and remeasured them in order to include 
realistic errors and incompleteness.

\subsection{Near-IR CMDs and luminosity functions}
\label{sect:near-ir_analysis}

\begin{figure}
\centering
\resizebox{\hsize}{!}{
\includegraphics[angle=0]{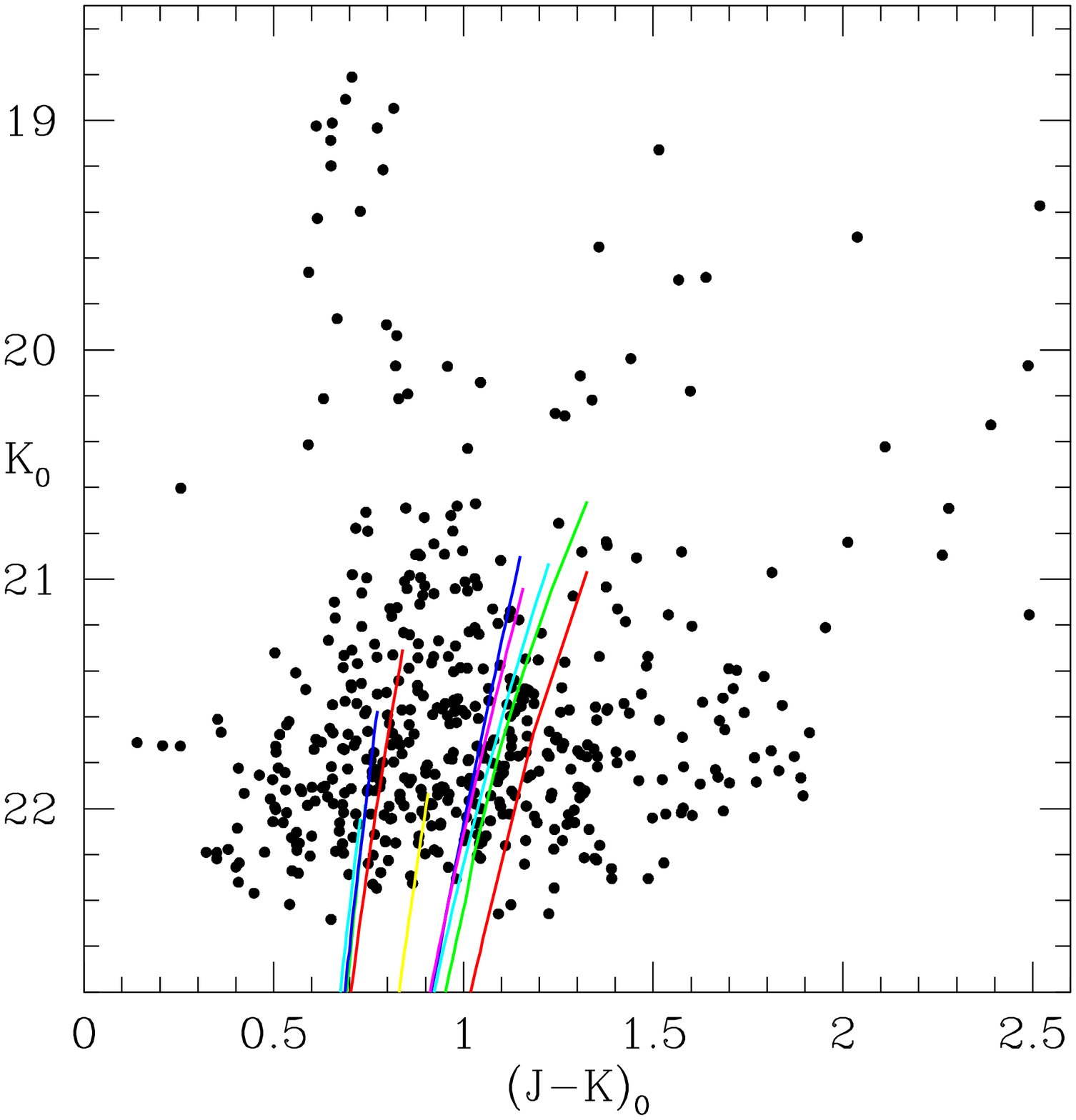}
}
\resizebox{\hsize}{!}{
\includegraphics[angle=0]{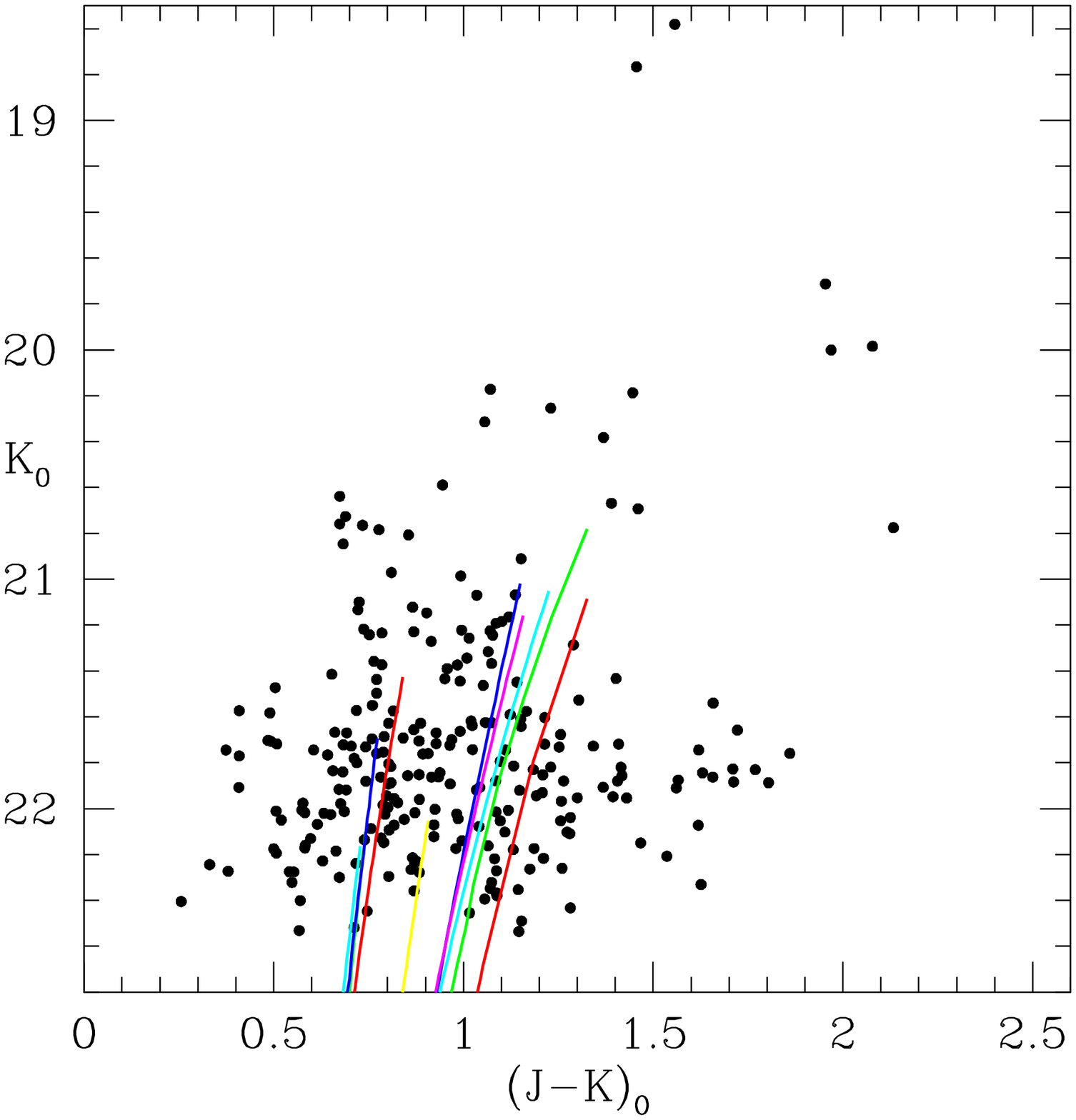}
}  
\caption[]
	{ISAAC $JK$ ``cleaned'' CMDs for AM~1339-445 (top panel) and AM~1343-452 
	(bottom) with overplotted fiducial red giant branches for the Galactic 
	globular clusters M15 ($\mathrm{[Fe/H]}=-2.17$), 
	M30 ($-2.13$), 
	M55 ($-1.81$), M4 ($-1.33$), M107 ($-0.99$), 47~Tuc ($-0.71$), M69 
	($-0.59$), NGC~6553 ($-0.29$), and NGC~6528 ($-0.23$)
	from \citet{ferraro+00}, corrected for the appropriate
	distance moduli and reddening, and shown from blue to red. 
	}
   \label{fig:CMDjkkclean}
\end{figure}

\begin{figure}
\centering
\resizebox{\hsize}{!}{
\includegraphics[angle=0]{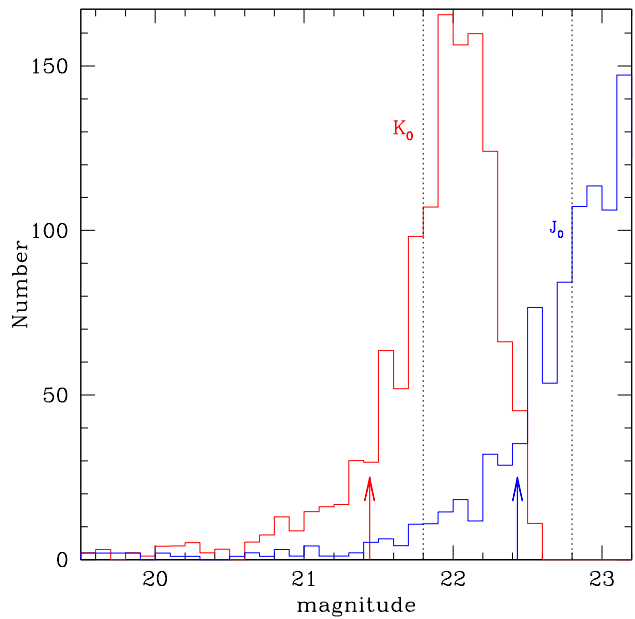}
}
\resizebox{\hsize}{!}{
\includegraphics[angle=0]{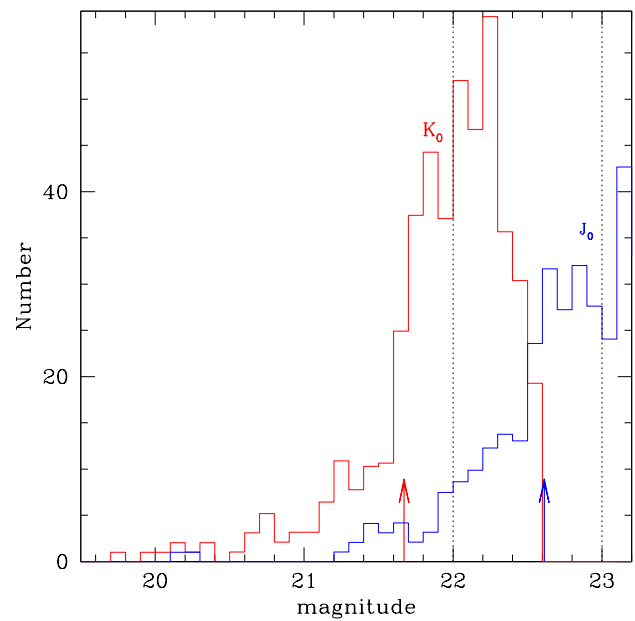}
}
\caption[]
	{Completeness corrected and field contamination 
	subtracted $J_0$ and $K_0$ luminosity
	functions for AM~1339-445 (top panel) and AM~1343-452 (bottom panel). 
	Dotted vertical lines indicate 50\% completeness
	limits, while the thick arrows show the expected magnitudes of the RGB tip in
	the two bands. For AM~1339-445 these are 
	$J_{0,TRGB}=22.43$ and $K_{0,TRGB}=21.44$, while for AM~1343-452 they 
	are $J_{0,TRGB}=22.61$ and $K_{0,TRGB}=21.67$.}
   \label{fig:LFjk}
\end{figure}

The statistically foreground and background decontaminated 
near-IR CMDs for both galaxies are shown in Fig.~\ref{fig:CMDjkkclean}. 
Due to larger size of the galaxy, and consequently smaller field area, 
the statistical decontamination worked less well in the case of AM~1339-445
leaving a plume of blue stars above the tip of the RGB. 
Overplotted on these CMDs 
are the RGB fiducials for Galactic globular clusters from the work of
\citet{ferraro+00}, which have been adjusted to the distance 
modulus adopted for each galaxy (see Sect.~\ref{sect:results_optical}).
The giant branches shown extend in metallicity from $\mathrm{[Fe/H]} = -2.2$ to
$\mathrm{[Fe/H]} = -0.2$. The large range of $(J-K)_0$ colours along the RGBs of 
these dE galaxies is almost entirely due to the photometric errors.
Unfortunately, incompleteness of around 50\% at magnitude $M_K=-5.5$ prevents 
us from determining reliably the average metallicity from these CMDs. We can 
only say that the average colour is not inconsistent with the mean metallicity 
calculated from the optical data.

In both galaxies only the brightest red giants are resolved. 
The field contamination subtracted and completeness corrected 
luminosity functions (LFs) for these bright giants are 
plotted in Fig.~\ref{fig:LFjk}.  
For both galaxies the short thick arrow indicates the expected location 
of tip of the 
RGB, calculated using the equations (7) and (9) from 
\citet{valenti+04b} and the distance moduli and mean abundances derived
from the optical data in the following section. 
For AM~1339-445 the expected RGB tip magnitudes are 
$J_{0,TRGB}=22.43 \pm 0.21$ and $K_{0,TRGB}=21.44 \pm 0.23$, 
while for AM~1343-452 they 
are $J_{0,TRGB}=22.61 \pm 0.21$ and $K_{0,TRGB}=21.67  \pm 0.23$. 
The 50\% incompleteness 
limit, plotted with dotted vertical lines, for both near-IR bands is only few 
tenths of a magnitude fainter than 
these expected RGB tip magnitudes, yet the LFs for both galaxies do show the
expected rise at or near the predicted RGB tip magnitudes, indicating 
consistency with the optical results.  In both galaxies, however, there
are a small number of stars above the RGB tip that could belong to an
intermediate-age AGB population. We investigate this possibility
in more detail in Sect.~\ref{sect:AGBcands} but we note here that at 
$K_{0} \approx 21.3$ and $J_{0} \approx 22.3$, the magnitude errors are 
$\pm 0.18$ and $\pm 0.19$, respectively for AM~1339-445, while the 
photometric errors for AM~1343-452 are few hundredths 
of magnitude smaller. Thus 
it is unlikely that this excess of stars above the RGB-tip results
entirely from magnitude errors for stars at the RGB-tip.

\subsection{Optical CMDs and luminosity functions}
\label{sect:results_optical}

\begin{figure}
\centering
\resizebox{\hsize}{!}{
\includegraphics[angle=270]{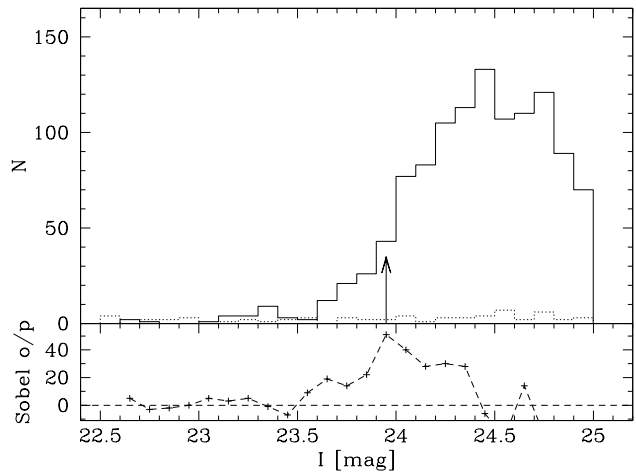}

}
\caption[]
	{Top panel: field subtracted (solid line) and field (dotted
line) $I$-band luminosity functions for AM~1339-445.  Bottom panel: output of
a Sobel filter applied to the field subtracted function.
	}
   \label{fig:LF_optical1}
\end{figure}

\begin{figure}
\centering
\resizebox{\hsize}{!}{

\includegraphics[angle=270]{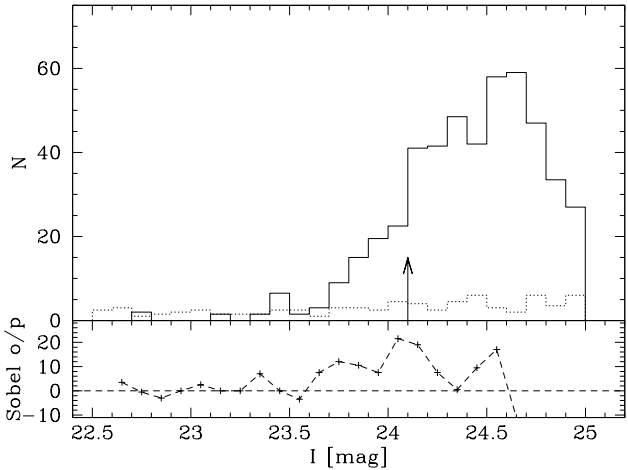}
}
\caption[]
	{Top panel: field subtracted (solid line) and field (dotted
line) $I$-band luminosity functions for AM~1343-452.  Bottom panel: output of
a Sobel filter applied to the field subtracted function.
	}
   \label{fig:LF_optical2}
\end{figure}

Figures \ref{fig:LF_optical1} and \ref{fig:LF_optical2} show the 
$I$-band LFs for AM~1339-445 and AM~1343-452, respectively.
In the upper panel of each figure, the dotted line shows the field
LF\@.  In both cases
it is possible that there are some galaxy stars in these adopted field
regions, but their numbers are likely to be small.  The solid curves in the
upper panels of the figures then show the field subtracted LFs for the
stars on the WF3 chips.  In both cases the field subtracted LFs hover
around zero at brighter magnitudes indicating that the adopted field
LFs are not grossly in error.  

A sharp increase in the number of stars in the LFs is expected at the RGB 
tip, and to aid in the identification of that feature, we show in the lower 
panels of the Figures the output of a Sobel edge-detection filter applied 
to the LF in the upper panels.  We adopt the peak in this function as
outlining the location of the RGB tip.  For AM~1339-445 this results
in $I_{TRGB}$
= 23.95 $\pm$ 0.15 while for AM~1343-452 we adopt $I_{TRGB}$ = 24.10 $\pm$ 
0.15, where in each case the uncertainty represents the combination of 
uncertainty in the peak location, taken as $\pm 0.10$~mag (the bin width),
and in the photometry zeropoint.  These
values agree very well with 23.93 $\pm$ 0.22 and 24.11 $\pm$ 0.25,
respectively, listed by \citet{k+02}.    

\citet{DA90} give a calibration (their equation 3) of the bolometric 
magnitude of the tip of the red giant branch based on the distance 
scale of \citet{LDZ90}, which includes a dependence on abundance [Fe/H].  
\citet{DA90} also give a relation (their equation 2) for the bolometric 
correction to the $I$ magnitude for red giants as function of $(V-I)_{0}$ 
colour.  Together these relations yield a calibration of the absolute
$I$ magnitude of the RGB tip, M$_{I}$(TRGB), as a function of [Fe/H] and
the dereddened $(V-I)$ colour of the RGB tip.  Similarly, 
\citet[][ see also \cite{armandroff+93}]{caldwell+98} give a calibration of 
abundance [Fe/H], on the \citet{ZW84} scale, with $(V-I)_{0,-3.5}$, the 
dereddened colour of the RGB at M$_{I} = -3.5$.  Together these two 
relations allow the calculation of the mean abundance and distance 
modulus for each of the galaxies, given $I$(TRGB), $(V-I)$(TRGB) and 
the reddening.
 
Using the $I$(TRGB) values given above, the reddening values given in 
Sect.\ \ref{sect:O+R} and $(V-I)_{0, TRGB}$ values from the CMDs shown 
in Figs.~\ref{fig:HSTcmds1339} and \ref{fig:HSTcmds1343} 
(the values adopted
were $1.70 \pm 0.03$ and $1.59 \pm 0.03$, respectively), we derive distance
moduli for the two galaxies as 
$(\mathrm{m}-\mathrm{M})_0(\mathrm{AM1339-445})=27.74 \pm 0.20$, and 
$(\mathrm{m}-\mathrm{M})_0(\mathrm{AM1343-452})=27.86 \pm 0.20$.  The
uncertainties listed take into account uncertainties in
the zeropoints of the photometry, the uncertainty in the derived mean
abundances and the uncertainty in the location of the RGB tip.
These values are again very similar to the distance  moduli determined 
by \citet{k+02}, who give slightly larger values 
($27.77 \pm 0.21$ for AM~1339-445, 
and $27.92 \pm 0.25$ for AM~1343-452) resulting from their use of a 
slightly brighter 
value for M$_{I}$(TRGB).  They are also in very good agreement with measured
distances from surface brightness fluctuation method by \citet{JFB00}, 
who list distances corresponding to moduli of
$27.87 \pm 0.27$ for AM~1339-445 and $27.99 \pm 0.37$ for AM~1343-452.

\begin{figure}
\centering
\resizebox{\hsize}{!}{
\includegraphics[angle=0]{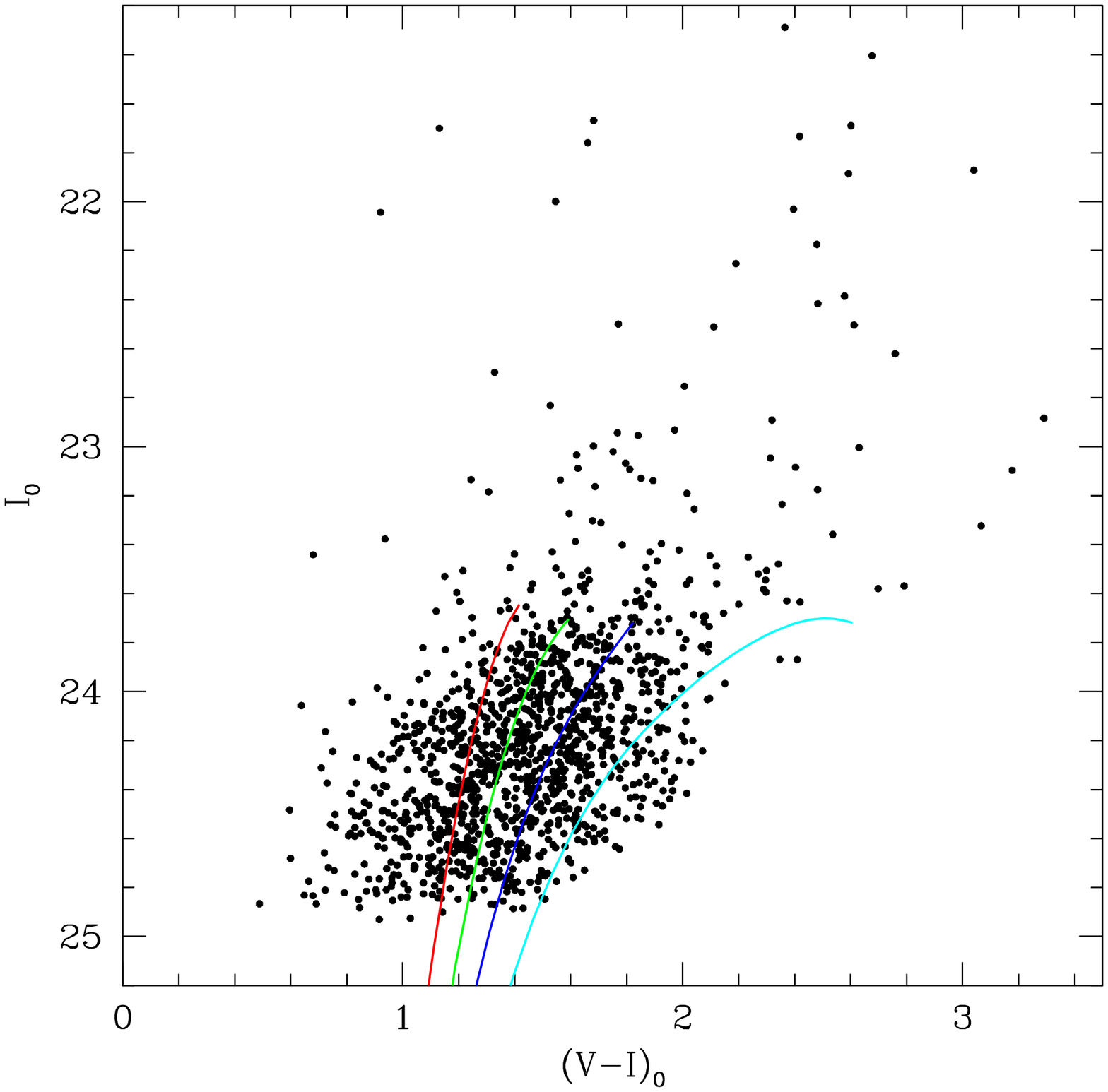}
}
\resizebox{\hsize}{!}{
\includegraphics[angle=0]{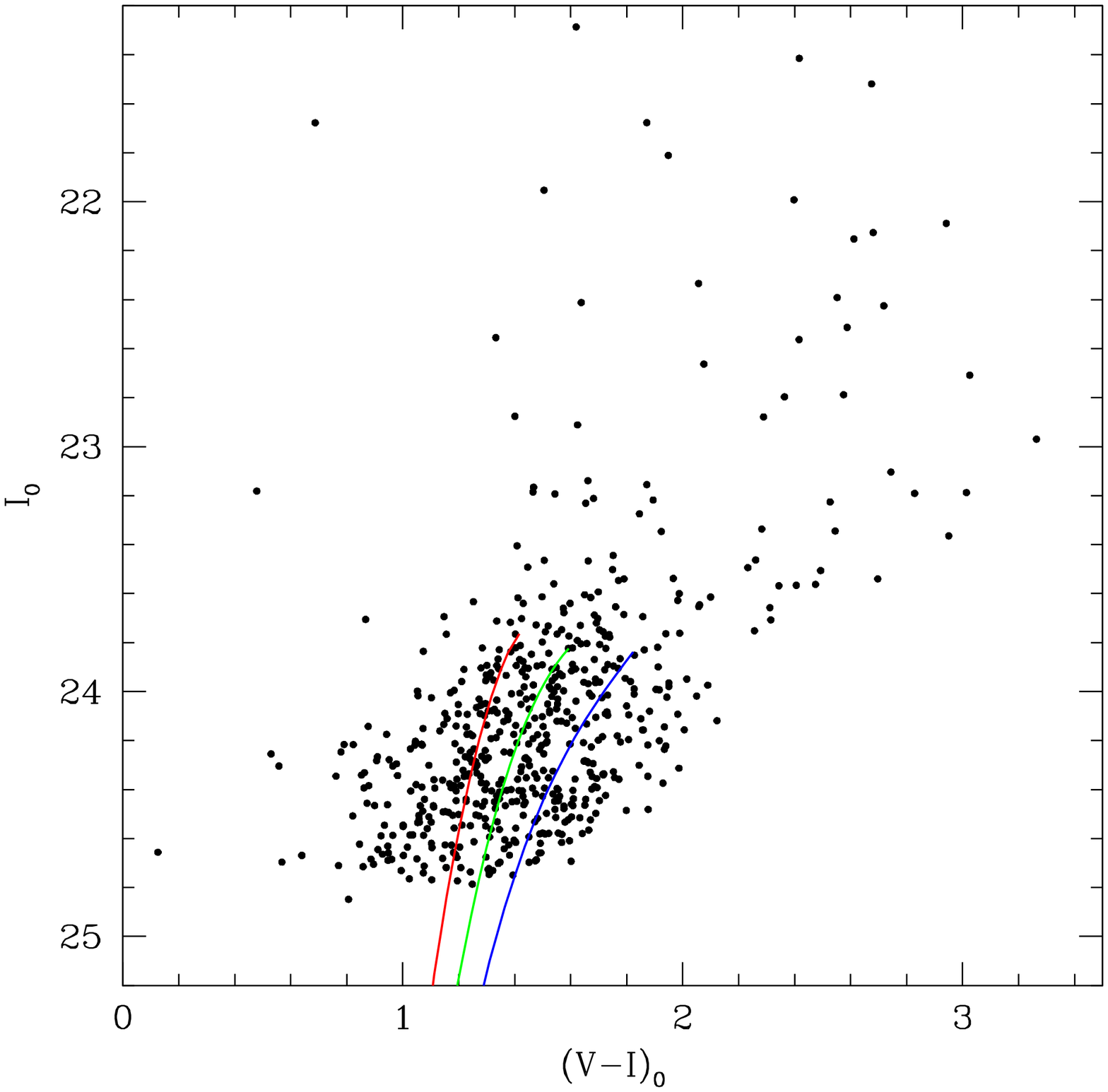}
}
\caption[]
	{WF3 $VI$ CMDs for AM~1339-445 (top panel) and 
	AM~1343-452 (bottom panel) with fiducial red giant
	branch for the Galactic globular clusters M15 ($\mathrm{[Fe/H]}
=-2.17$), M2 ($-1.62$), NGC~1851 ($-1.36$) and 47~Tuc ($-0.71$) overplotted,
using the appropriate distance moduli and reddening.  The 47~Tuc giant
branch is not plotted for AM~1343-452.
	}
   \label{fig:RGBsVI}
\end{figure}

The mean metallicities of the two galaxies,
as determined from the mean colour of the RGBs in a $\pm 0.1$ mag interval 
in $I$ about M$_{I}= -3.5$, are
$\langle \mathrm{[Fe/H]}\rangle =-1.4 \pm 0.2$ for AM~1339-445 and 
$\langle \mathrm{[Fe/H]}\rangle =-1.6 \pm 0.2$ for AM~1343-452.  
The listed uncertainty here includes the effect of uncertainty in
the distance moduli, the statistical and photometric zeropoint uncertainty, 
$\pm 0.02$ and $\pm 0.03$~mag, respectively,
and uncertainty in the abundance calibration. These four contributions 
are all approximately equal in size and have been added in quadrature to 
determine the overall uncertainty. In Fig.~\ref{fig:RGBsVI} we show the WF3
photometry for each galaxy with the giant branches of standard globular
clusters from \citet{DA90} overplotted, using the
distance moduli derived above.  The four clusters are M15 ($\mathrm{[Fe/H]}
=-2.17$), M2 ($-1.62$), NGC~1851 ($-1.36$) and 47~Tuc ($-0.71$). We also note
that, due to the limited exposure times and the consequent size
of the errors
in the colours for the individual RGB stars, it is not possible to place any
meaningful constraints on the size of any metallicity spread 
in these galaxies.  

Strictly speaking the average metallicity derived in this way is 
only valid for stellar populations with an age comparable to that of 
Galactic globular clusters. \citet{saviane+00} discuss the effect of 
the younger mean age of the population and the RGB colour in Fornax dE. 
They find that the bluer colours of the 5 Gyr old population mimic a 
$\sim 0.4$~dex more metal-poor stellar population of 15~Gyr. Later we 
will argue that a significant intermediate-age population is
present in our targets, although with a much lower fraction than in Fornax. 
Hence we estimate that our average metallicities might be systematically 
too low by up to $\sim 0.2$~dex. Fortunately, none of the conclusions in the
following sections are significantly altered if the true metallicities exceed
those listed by amounts of this order.
A more detailed discussion of the influence of the composite stellar 
populations on the photometric metallicities and on the measurement of
distances via RGB tip method is available in \citet{salaris+girardi05}.

In general, the optical CMDs (Fig.~\ref{fig:RGBsVI}) 
are consistent with the old metal-poor population
expected for dE galaxies.  They are dominated by red giants and there
is no indication of any (young) blue stars.  The relatively low Galactic
latitude of the fields, however, results in significant contamination of 
the CMDs by foreground Galactic stars, particularly for magnitudes above the 
red giant branch tip. Most of them are field red clump stars that have 
optical colours similar to metal-poor RGB and AGB stars. Identification 
of any upper-AGB stars is therefore
not as straightforward as it is for dEs in higher latitude fields
\citep[cf.][]{caldwell+98}.

The LFs in Figs.~\ref{fig:LF_optical1} and 
\ref{fig:LF_optical2} clearly show an apparent excess 
of stars above the field LFs at luminosities above that of the RGB tip.
At the tip of the RGB, the typical error in the $I$ magnitudes is
$\sigma(I) \approx 0.1$ mag.  Consequently, given that any star 
whose photometry was affected by crowding by more than 0.05 mag has
been excluded, it is entirely plausible that these stars, particularly those
more than 0.2-0.3 mag above the RGB tip, are indeed intermediate-age 
upper-AGB stars.
We now use our combined optical/near-IR dataset to investigate this
possibility. 

\subsection{Candidate upper-AGB stars}
\label{sect:AGBcands}

\begin{table*}
\caption{Dereddened magnitudes in $V$, $I$, $J_s$ and the two measurements 
in the $K_s$ filter of AGB star candidates. 
In the last two columns we list absolute 
bolometric magnitudes of these stars calculated using the appropriate 
distance modulus and $[BC_{I_c},(V-I)_c]$ relation of \citet{DA90}, 
and $[BC_K,(J-K)]$ of \citet{costa+frogel96}, in columns 9 and 10, respectively.
}
\label{tab:AGBstars}
\centering
\begin{tabular}{lccccccccc}
\hline
\hline
ID           & X(WF3)  & Y(WF3) &$V_0(\pm \sigma)$&$I_0(\pm \sigma)$& $J_0(\pm \sigma)$ & $K_1(\pm \sigma)$ & $K_2(\pm \sigma)$ &$\mathrm{M}_{bol}(VI)$&$\mathrm{M}_{bol}(JK)$\\
\hline
\multicolumn{8}{c}{AM~1339-445}\\
\hline
 560   & 352.55  & 527.38 & 25.591 (0.109) &23.236 (0.066) &22.35 (0.08) &21.00 (0.09) & 21.21 (0.16)  &  $-$4.20 &$-$3.75\\
 684   & 367.45  & 444.05 & 25.186 (0.083) &23.402 (0.070) &21.87 (0.06) &20.93 (0.10) & 20.84 (0.07)  &  $-$3.89 &$-$4.24\\
 720   & 325.04  & 378.23 & 24.904 (0.066) &22.933 (0.051) &21.70 (0.05) &20.73 (0.08) & 20.61 (0.07)  &  $-$4.41 &$-$4.40\\
 729   & 119.52  & 194.34 & 25.361 (0.084) &23.047 (0.053) &21.76 (0.06) &20.82 (0.09) & 20.73 (0.11)  &  $-$4.38 &$-$4.35\\
 780   & 499.28  & 480.96 & 24.981 (0.072) &23.303 (0.073) &22.27 (0.09) &21.28 (0.13) & 21.31 (0.15)  &  $-$3.97 &$-$3.84\\
 812   & 130.06  & 140.34 & 24.796 (0.059) &22.955 (0.050) &21.51 (0.05) &20.41 (0.07) & 20.30 (0.07)  &  $-$4.36 &$-$4.58\\
 817   & 576.48  & 522.78 & 24.981 (0.075) &23.130 (0.059) &21.44 (0.04) &20.48 (0.06) & 20.34 (0.07)  &  $-$4.18 &$-$4.67\\
 857   & 351.16  & 300.80 & 25.033 (0.071) &23.139 (0.060) &21.78 (0.06) &20.20 (0.05) & 20.16 (0.05)  &  $-$4.19 &$-$4.39\\
 1046  & 501.73  & 276.17 & 24.904 (0.075) &23.093 (0.057) &21.52 (0.05) &20.28 (0.06) & 20.27 (0.06)  &  $-$4.21 &$-$4.57\\
 1087  & 389.85  & 153.30 & 25.297 (0.084) &23.256 (0.062) &22.21 (0.08) &20.81 (0.08) & 20.92 (0.15)  &  $-$4.10 &$-$3.90\\
 1114  & 761.45  & 447.66 & 25.019 (0.073) &23.311 (0.066) &22.35 (0.09) &21.21 (0.12) & 20.92 (0.09)  &  $-$3.96 &$-$3.74\\
\hline
\multicolumn{8}{c}{AM~1343-452}\\
\hline
 890   &  219.26 & 627.35 & 26.237 (0.194) &23.540 (0.081) &22.38 (0.10) &21.40 (0.14) & 21.22 (0.13) &  $-$4.10  &$-$3.84 \\
 982   &  409.25 & 691.57 & 25.318 (0.095) &23.547 (0.084) &22.39 (0.08) &21.35 (0.14) & 21.52 (0.16) &  $-$3.87  &$-$3.86 \\
 1042  &  485.46 & 695.75 & 26.020 (0.157) &23.191 (0.062) &21.37 (0.04) &20.27 (0.05) & 20.35 (0.05) &  $-$4.48  &$-$4.85 \\
 1197  &  581.90 & 657.49 & 25.999 (0.160) &23.506 (0.080) &22.06 (0.08) &20.62 (0.05) & 20.79 (0.08) &  $-$4.09  &$-$4.17 \\
 1232  &  484.85 & 583.75 & 24.536 (0.057) &22.911 (0.054) &22.43 (0.12) &20.71 (0.06) & 20.90 (0.08) &  $-$4.47  &$-$3.88 \\
 1484  &  350.15 & 394.85 & 25.113 (0.084) &23.218 (0.067) &21.97 (0.11) &20.34 (0.07) & 19.98 (0.04) &  $-$4.23  &$-$4.52 \\
 1526  &  342.88 & 370.90 & 25.270 (0.093) &23.346 (0.073) &21.53 (0.05) &20.63 (0.06) & 20.34 (0.06) &  $-$4.11  &$-$4.71 \\
 1542  &  356.72 & 373.95 & 25.619 (0.135) &23.336 (0.072) &21.75 (0.07) &20.88 (0.10) & 20.72 (0.09) &  $-$4.21  &$-$4.49 \\
 1587  &  314.38 & 326.92 & 25.161 (0.082) &22.797 (0.049) &21.79 (0.05) &20.49 (0.06) & 20.53 (0.06) &  $-$4.76  &$-$4.42 \\
 1572  &  293.25 & 322.81 & 25.317 (0.100) &23.765 (0.096) &22.22 (0.08) &21.29 (0.10) & 20.81 (0.09) &  $-$3.59  &$-$4.01 \\  
\hline												      		   
\end{tabular}	   									    
\end{table*}				   									    

\begin{figure}
\centering
\resizebox{\hsize}{!}{
\includegraphics[angle=0]{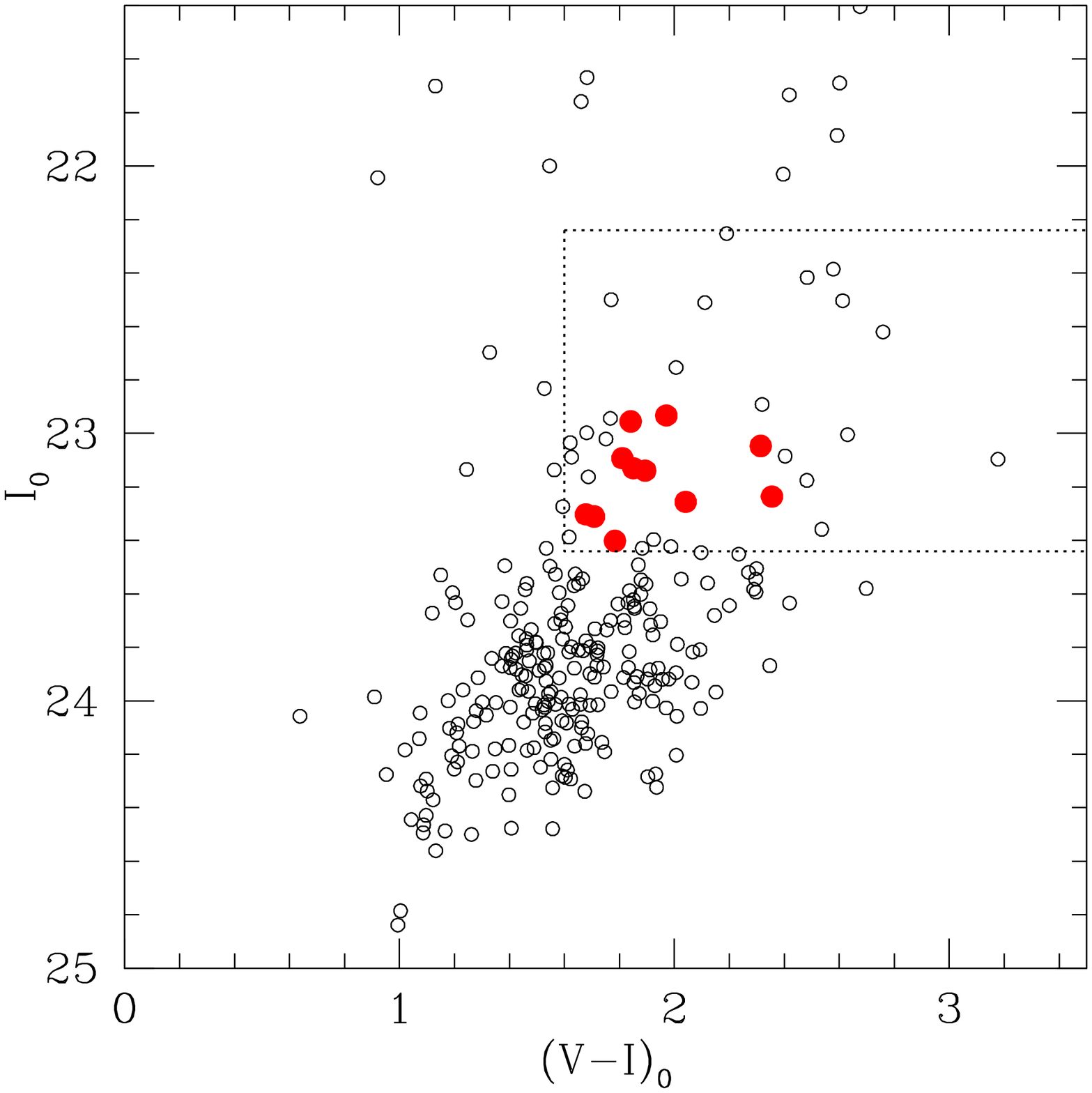}
\includegraphics[angle=0]{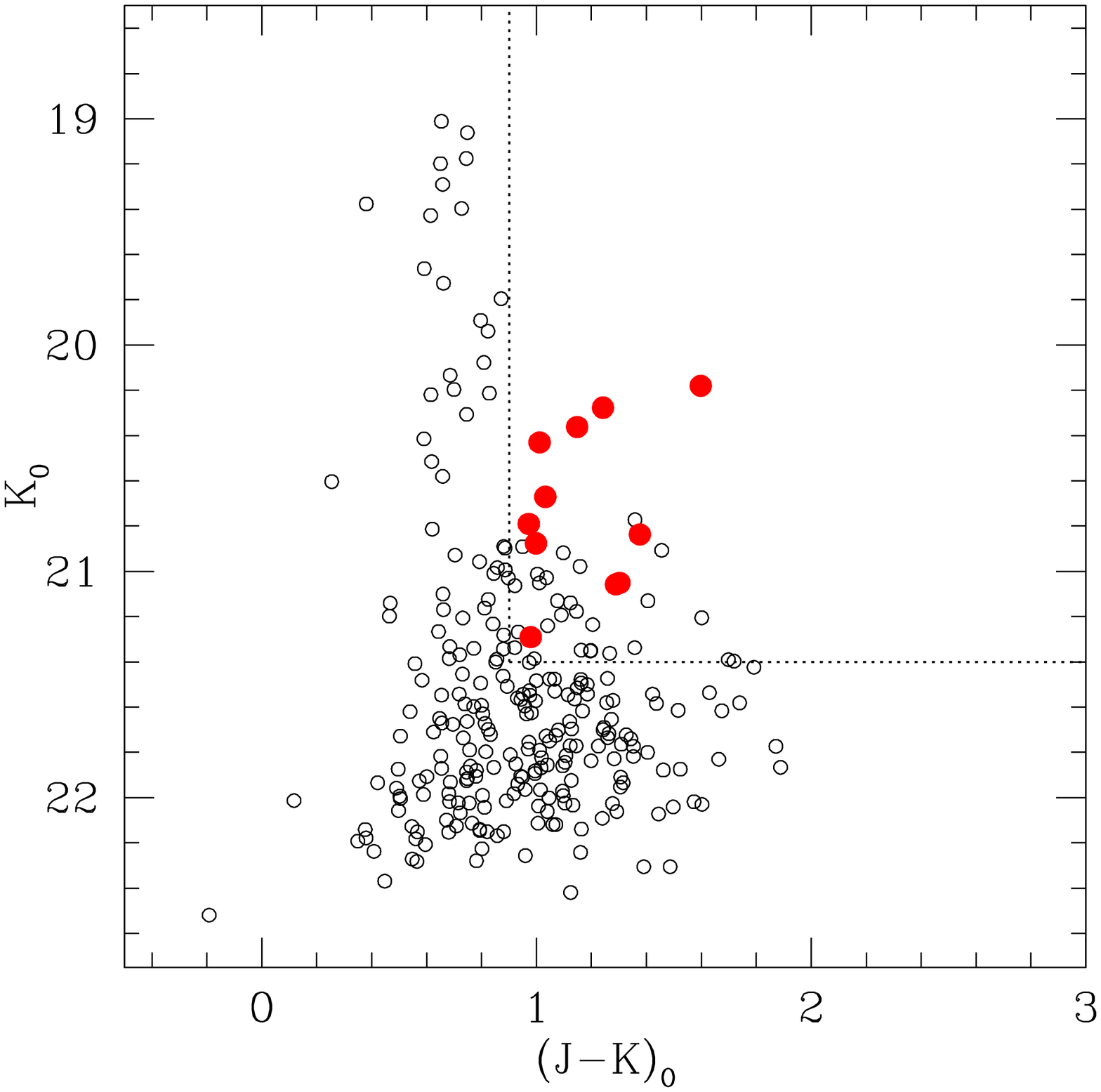}
}
\caption[]
	{$VI$ and $JK$ CMDs of AM~1339-445 with adopted box for the selection of 
	upper-AGB candidates. All the candidate upper-AGB stars are plotted with
	large filled symbols. 
	}
   \label{fig:am1339_agb_allsel}
\end{figure}

\begin{figure}
\centering
\resizebox{\hsize}{!}{
\includegraphics[angle=0]{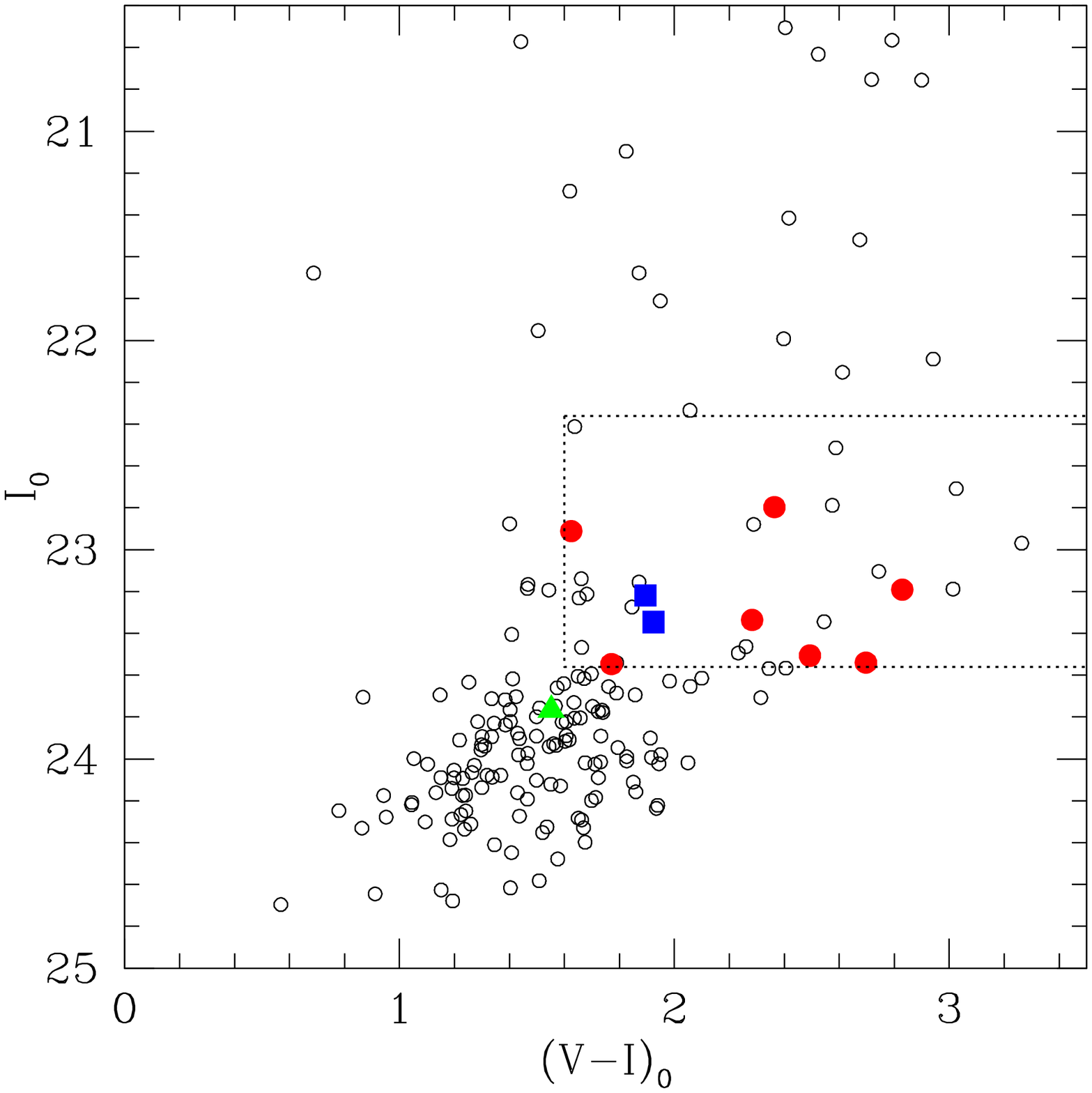}
\includegraphics[angle=0]{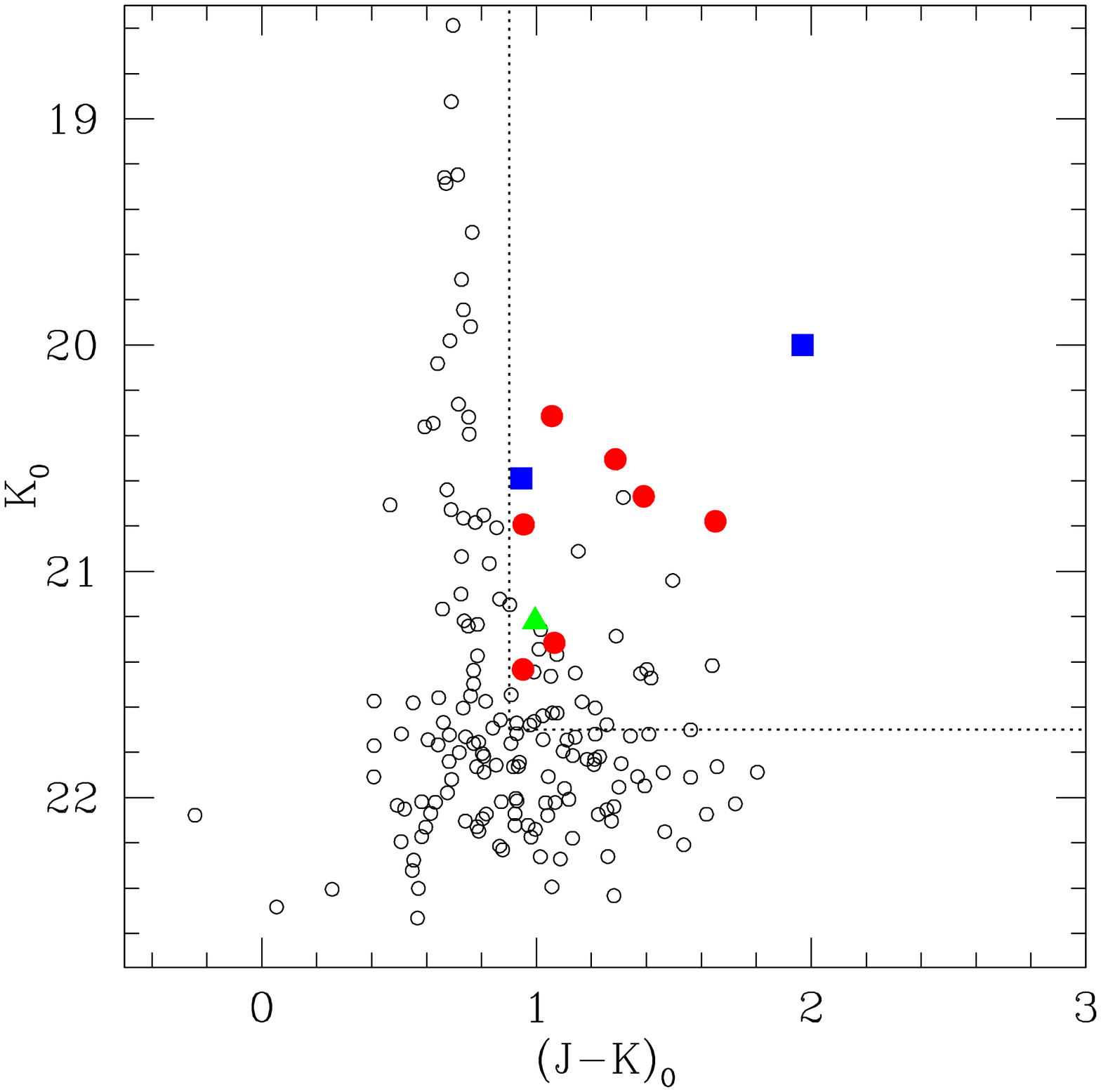}
}
\caption[]
	{$VI$ and $JK$ CMDs of AM~1343-452 with adopted box for the selection of 
	upper-AGB candidates. All the candidate upper-AGB stars are plotted with
	large filled symbols. The three candidate long period variable stars are
	plotted as large (blue) squares and a (green) triangle.}
   \label{fig:am1343_agb_allsel}
\end{figure}

Asymptotic giant branch stars (AGB) brighter than the RGB tip are found in 
relatively metal-poor intermediate-age populations, and in metal-rich
($\mathrm{[Fe/H]}\ga -1$~dex) and old systems \citep{frogel+whitelock98,
guarnieri+98}. 
In metal-poor old Galactic globular clusters, stars brighter than the RGB tip 
are not observed, while they are present in the intermediate-age LMC 
and SMC clusters
\citep{frogel+90}. Since the estimated average metallicity of our target 
galaxies shows that they are relatively poor in metals, any bright AGB stars 
above the tip of the RGB are expected to belong to an intermediate-age 
population. Due to their cool atmospheres these stars are particularly
bright in the near-IR wavelength range. 
Moreover, they are typically redder than Galactic foreground stars,
and thus relatively easy to detect. In optical CMDs, however, 
this distinction is less easy.
The availability of matched optical and near-IR photometry allows
us to investigate the presence or absence of bright AGB stars in our target 
galaxies. 

In Sect.~\ref{sect:ISAAC-HST-combine} we described the combination of
the optical and near-IR datasets to produce a sample of stars in both
galaxies with $VI$ and $JK$ photometry.  The CMDs for these combined
datasets are shown in Figures~\ref{fig:am1339_agb_allsel} and 
\ref{fig:am1343_agb_allsel} for AM~1339-445 and AM~1343-452, respectively.
We have adopted the following criteria to identify candidate upper-AGB stars
in these diagrams.  First, in the near-IR CMDs we insist that any upper-AGB
star candidate be redder than the field sequence and be brighter than the 
RGB tip.  In other words, the candidates must satisfy $(J-K)_{0}$ $>$ 0.90
and $K_{0}$ $<$ 21.4 (AM~1339-445) or $K_{0}$ $<$ 21.7 (AM~1343-452).  These
boundaries are shown as the dotted lines in the $JK$ CMDs.  To place 
similar limits in the $VI$ CMDs we use as guide the HST WFPC2 imaging of two
M81 group dEs discussed in \citet{caldwell+98}.  Although the two dwarfs 
studied,
F8D1 and BK~5N, are at a comparable distance to the dEs
studied here, the considerably longer total exposure times and the
comparatively high Galactic latitude of the M81 group, results in precise
CMDs that are essentially free of foreground and background contamination.
Both these M81 group dEs show clear upper-AGB populations \citep{caldwell+98},
and we use these populations to guide our selection of candidates
in the $VI$ CMDs for the Cen~A group dEs. We note that the
upper-AGB populations occupy a region in the M81 group dEs
 corresponding to 1.6 $\leq$ 
$(V-I)_{0} \leq 3.3$ and $-4.3 \geq \mathrm{M}_{I} \geq -5.5$.  Here
the fainter limit has been specifically chosen to exceed the RGB tip
magnitude by $\sim$0.3 to minimise, for the Cen~A group dEs where the 
uncertainties are larger, the possibility of false candidates
arising from RGB tip stars displaced upwards by photometry errors.  This
selection region is outlined by the dotted-line rectangles in the $VI$
CMDs.

The stars that satisfy the criteria for candidate upper-AGB stars
in {\it both} the $VI$ and $JK$ CMDs are plotted with filled symbols in
Figures~\ref{fig:am1339_agb_allsel} and \ref{fig:am1343_agb_allsel}.
There are 11 such stars in AM~1339-445 and 9 in AM~1343-452. 
The two long period variable star candidates in AM~1343-452 (\#1484 and \#1526; 
see below) that satisfy both selection criteria in $VI$ and $JK$ CMDs are 
plotted with filled (blue) squares. In addition, one star with an $I$ magnitude
only slightly brighter than the RGB tip (\#1572; shown with large filled (green)
triangle in the CMDs), and thus fainter than our conservative limit set to
$0.3$~mag brighter than the RGB tip in the $I$ filter, 
was included in the list of candidate AGB stars in AM~1343-452 
as it is also a potential long period variable star. The full set of 
photometry measures for these candidate
upper-AGB stars are given in Table~\ref{tab:AGBstars}.  

\begin{figure}
\centering
\resizebox{\hsize}{!}{
\includegraphics[angle=0]{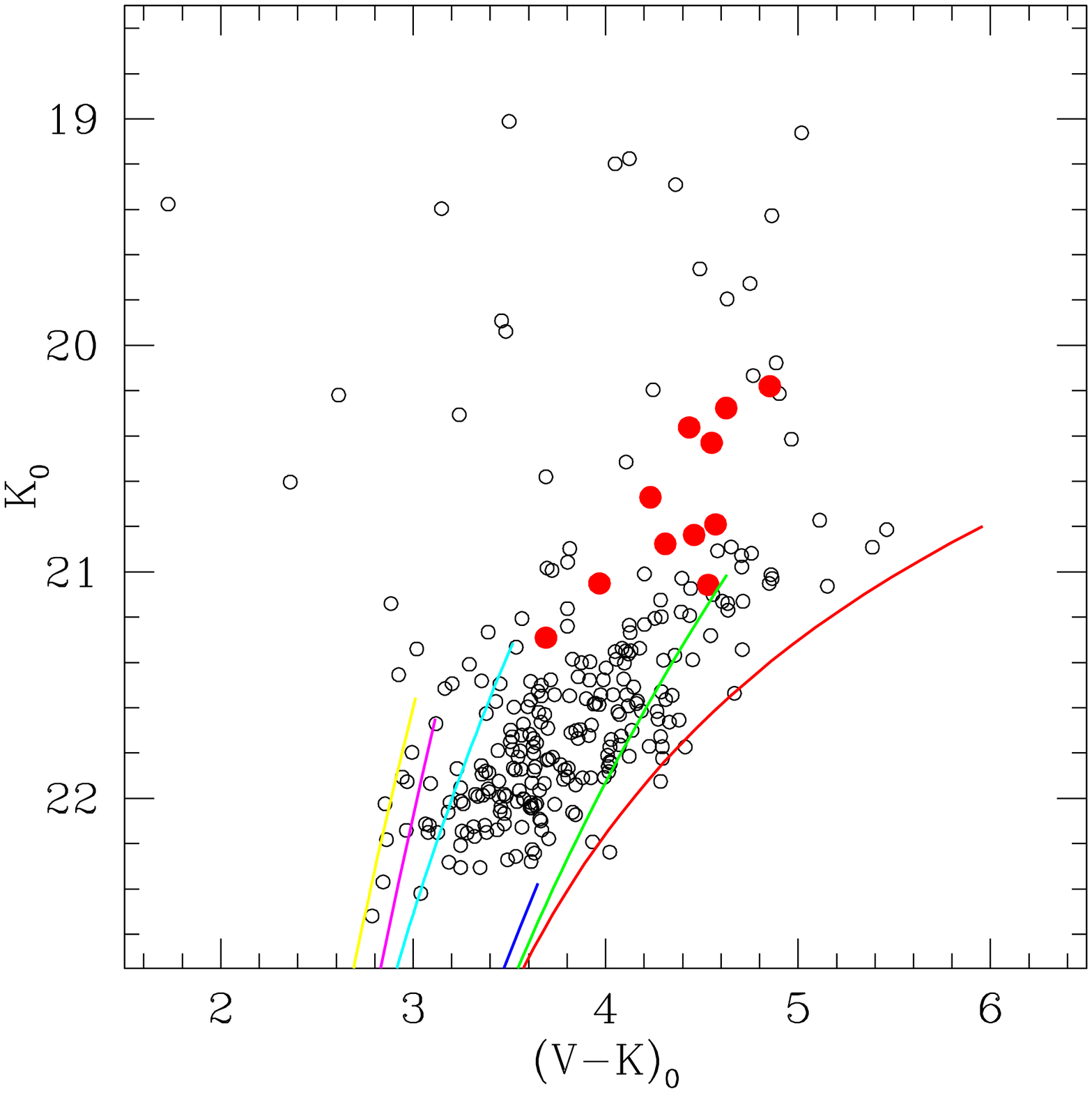}
\includegraphics[angle=0]{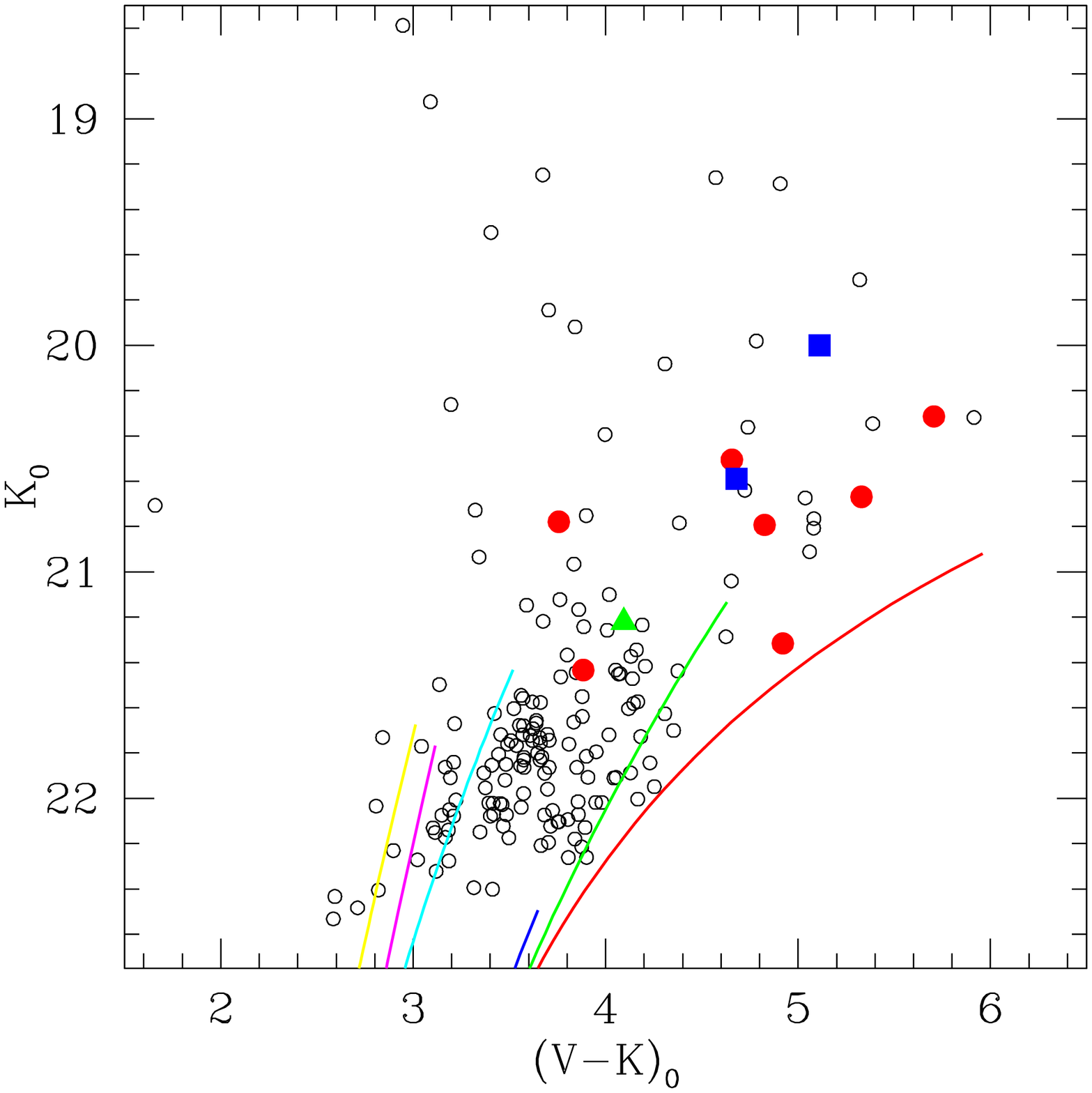}
}
\caption[]
	{$VK$ CMDs for the stars found in common in HST WF3 and ISAAC
	$J_s$ and $K_s$ images for AM~1339-445 (left panel) and AM~1343-452 (right
	panel). The lines are Galactic globular cluster RGB fiducials for
	M15($\mathrm{[Fe/H]}=-2.17$), M30 ($-2.13$), M55 ($-1.81$), M4 ($-1.33$), 
	M107 ($-0.99$) and 47~Tuc ($-0.71$) from \cite{ferraro+00}, plotted using 
	the appropriate distance moduli and reddening, and going from blue to red.
	The large filled symbols 
	show the position of upper-AGB star candidates in these two galaxies. 
	The two candidate LPVs in AM~1343-452 that are located within the 
	selection boxes in Fig.~\ref{fig:am1343_agb_allsel} are plotted with 
	filled (blue) squares and the additional LPV candidate with slightly 
	fainter $I$-band magnitude than our selection limit is plotted with a
	(green) triangle.  
	}
   \label{fig:vkcmds_agb}
\end{figure}

In Fig.~\ref{fig:vkcmds_agb} we show $(V-K)_{0}$ vs.\ $K_{0}$ CMDs for both 
galaxies with 
RGB fiducials for Galactic globular clusters M15($\mathrm{[Fe/H]}=-2.17$), 
M30 ($-2.13$), M55 ($-1.81$), M4 ($-1.33$), M107 ($-0.99$) and 47~Tuc ($-0.71$) 
from \cite{ferraro+00} overplotted using the appropriate distance 
moduli and reddening. As can be seen by the wide spread in colour of these 
globular
cluster RGB fiducials, $(V-K)_0$ colour is particularly sensitive to metallicity. 
The $(V-K)_0$ colours of the red giants in these two galaxies are bracketed by
M55 and M107 ridge lines as expected from the average metallicity derived 
from the
optical data alone. The AGB candidates are plotted with larger filled symbols 
and are seen to have magnitudes 
brighter than the RGB tip and relatively red $(V-K)$ colours.  This is further
confirmation that these stars are plausible upper-AGB candidates.
\begin{figure}
\centering
\resizebox{\hsize}{!}{
\includegraphics[angle=270]{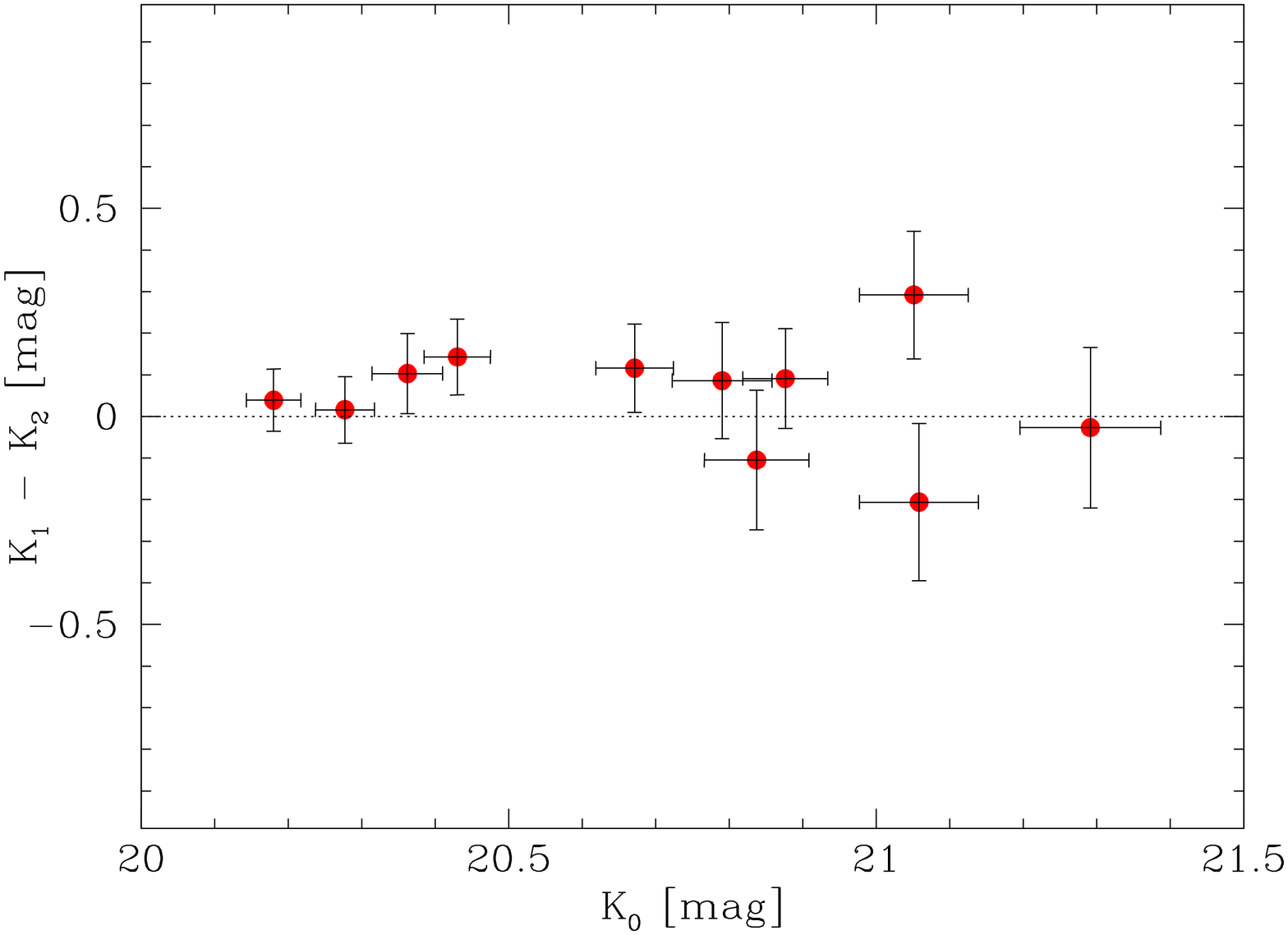}
}
\resizebox{\hsize}{!}{
\includegraphics[angle=270]{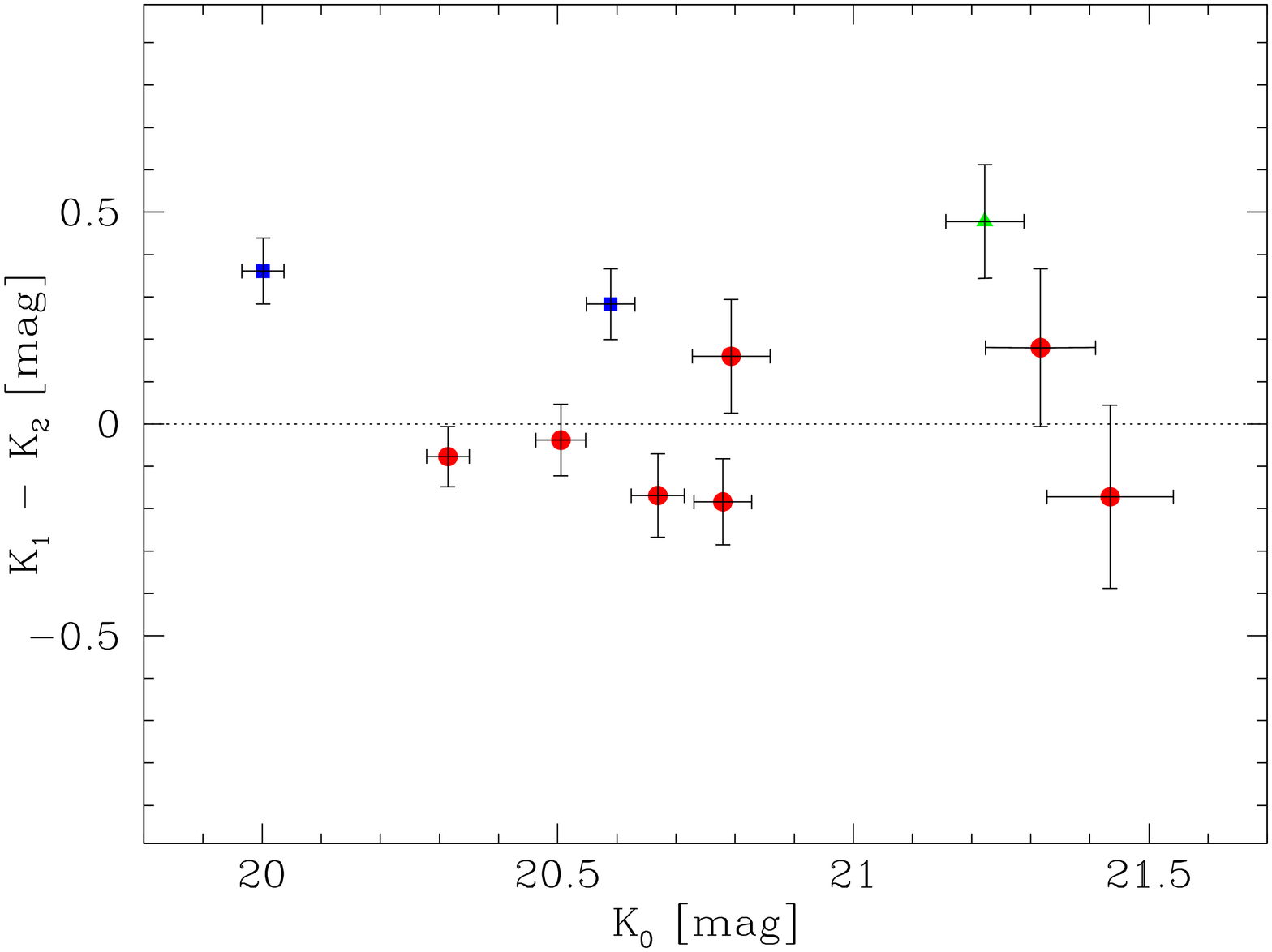}
}
\caption[]
	{Top panel: $K$-band magnitude difference measured between the
two epochs for the upper-AGB star candidates in AM~1339-445.  Due to the only 
4 day interval between the two observations, none of the stars displays 
any variability.
Bottom panel: $K$-band magnitude difference measured between the
two epochs for the upper-AGB star candidates in AM~1343-452. The time 
interval between the two epochs is 43 days. Three of the stars, plotted as 
filled (blue) squares and filled (green) triangle (see text), are LPV 
candidates as their magnitudes
vary by more than 0.25~mag and the difference is more than $3.3~\sigma$ larger
than the combined photometric errors.}
   \label{fig:am_agbvar}
\end{figure}

\begin{table}
\caption{K-band photometry and positions of two very red objects in field 
of AM~1343-452.}
\label{tab:redstars}
\centering
\begin{tabular}{lcccc}
\hline
\hline
ID           & $\alpha_{2000}$  & $\delta_{2000}$ & $K_1(\pm \sigma)$ & $K_2(\pm \sigma)$ \\
\hline
1182  & 13:46:13 & $-$45:41:27  &21.09 (0.11) & 21.12 (0.12) \\
2040  & 13:46:15 & $-$45:40:52  &20.81 (0.10) & 20.61 (0.08) \\
\hline		
\end{tabular}
\end{table}	   	

We also searched for stars with very red $(J_s - K_s)$ colours, stars 
that are well measured on both K images but which have no 
counterpart on the J-band image. Such stars might be dust enshrouded AGB stars,
similar to those present in Leo~I \citep{menzies+02} and F8D1 
\citep{dacosta04}.
Two such very red stellar objects are indeed present in AM~1343-452, 
while none were found in AM~1339-445. Neither of the two very red stars in 
AM~1343-452 show significant variability between the two K magnitude 
measurements. 
One star (\#2040) is within the field of view of WF3 chip of WFPC2, 
but is not detected in 
the F814W image. The second red object (\#1182) is found well outside the
main body of the galaxy, and just outside of the WFPC2 field of view. 
While it is possible that that object is a compact high
redshift galaxy, it is also not excluded that it belongs to AM~1343-452. For
example, in Leo~I one of the five very red stars is also found well away 
from the center of its host galaxy 
\citep{menzies+02}. The photometry and positions of the two very red stars in 
the field of AM~1343-452 are listed in Table~\ref{tab:redstars}.  Given that
for these stars $J_{0}$ exceeds 23, the 50\% completeness limit, they must 
have $(J-K)_0$ $\geq$ 1.9 and 2.3, respectively.

A distinguishing characteristic of upper-AGB stars is that most, 
if not all, are long
period variables (LPV) displaying large amplitude variations, particularly
at optical wavelengths \citep[e.g.][]{whitelock+00}.  
For example, a large fraction of the upper-AGB stars
identified by \citet{caldwell+98} in the M81 group dEs are evidently 
variables 
\citep{caldwell+98}.
As the name implies, the periods for the variability are long, typically 
hundreds of days.  These amplitude and period properties make such stars
relatively easy to detect, given sufficient epochs, from the ground even 
in crowded fields and at distances of 4~Mpc or larger \citep{rejkuba+03}.
Variability is thus a further criterion against which we can assess our
candidate upper-AGB stars.  Unfortunately, due to an operational oversight, 
the two $K$-band epochs for AM~1339-445 were observed only 4 days apart.  
This is much smaller 
than the variability timescale of LPVs and thus it is not surprising that 
none of the upper-AGB candidates display variability (Fig.~\ref{fig:am_agbvar} 
top panel).
On the other hand, the AM~1343-452 $K$-band observations are separated by 
43 days.  Of course, with just two epochs we can only set a lower limit 
to the number of variables, as some may change very little in the interval 
or be caught at two phases that have similar brightnesses. 
Nevertheless it is encouraging to see in the bottom panel 
of Fig.~\ref{fig:am_agbvar}, where we plot $\Delta K = K_1 - K_2$, the 
$K$-band magnitude difference measured between the
two epochs for the AM~1343-452 upper-AGB star candidates, that two of these 
stars, \#1484 and \#1526 (filled blue squares), 
are indeed probable variables.  
In both cases the magnitude difference
is more than $3.3~\sigma$ larger than the combined photometric
errors of the two measurements. This result suggests that we are indeed
selecting genuine candidate upper-AGB stars. Based on such variability 
criteria, 
we add one more star, \#1572, to the list of upper-AGB candidates. 
This one
has $I$ magnitude of 23.77 and it is only very slightly brighter than the RGB 
tip
in AM~1343-452 ($I_{0,TRGB}=23.86$), and thus outside our conservative 
AGB selection 
box in the $VI$ CMD, while it is within the AGB selection box in the $JK$ CMD. 
We plot
this star with filled (green) triangle in both panels in
Fig.~\ref{fig:am1343_agb_allsel}.  With this star and the two very red objects,
our final upper-AGB candidate list for AM~1343-452 contains a total of
12 stars, while that for AM~1339-445 remains at 11.  We will discuss the 
characteristics of these stars and the implications they provide for the
star formation histories of the dEs in Sect.~\ref{sect:discuss}.

%
%

\subsection{Globular Clusters}

\begin{table*}
\caption{Photometry of Globular Cluster Candidates in AM~1339-445. The first
three columns are the identifier, and the coordinates of the candidate globular
cluster from \citet{sharina+05}. After that we list their 
(X,Y) positions in the ISAAC $J_s$ band image and integrated apparent 
V, I, J and K magnitudes (corrected for Galactic extinction using
\citet{schlegel+98} maps) and corresponding errors.}
\label{tab:GCs}
\centering
\begin{tabular}{lccccllll}
\hline
\hline
ID    & $\alpha_{2000}$ & $\delta_{2000}$ & X(ISAAC) & Y(ISAAC) &
$V_0$ &  $I_0$          & $J_0$         & $K_0$      \\
\hline
KK211--3--917&13:42:08.0 & $-$45:12:29 &  611.21 & 500.09 & 
$20.91 \pm 0.07$ & $20.00 \pm 0.07$ & $19.3 \pm 0.15$ & $18.76 \pm 0.10$ \\
KK211--3--149&13:42:05.6 & $-$45:12:20 &  435.92 & 433.74 & 
$19.95 \pm 0.07$ & $19.00 \pm 0.07$ & $18.4 \pm 0.15$ & $17.74 \pm 0.10$ \\
\hline
\end{tabular}
\end{table*}

\begin{figure}
\centering
\resizebox{\hsize}{!}{
\includegraphics[angle=0]{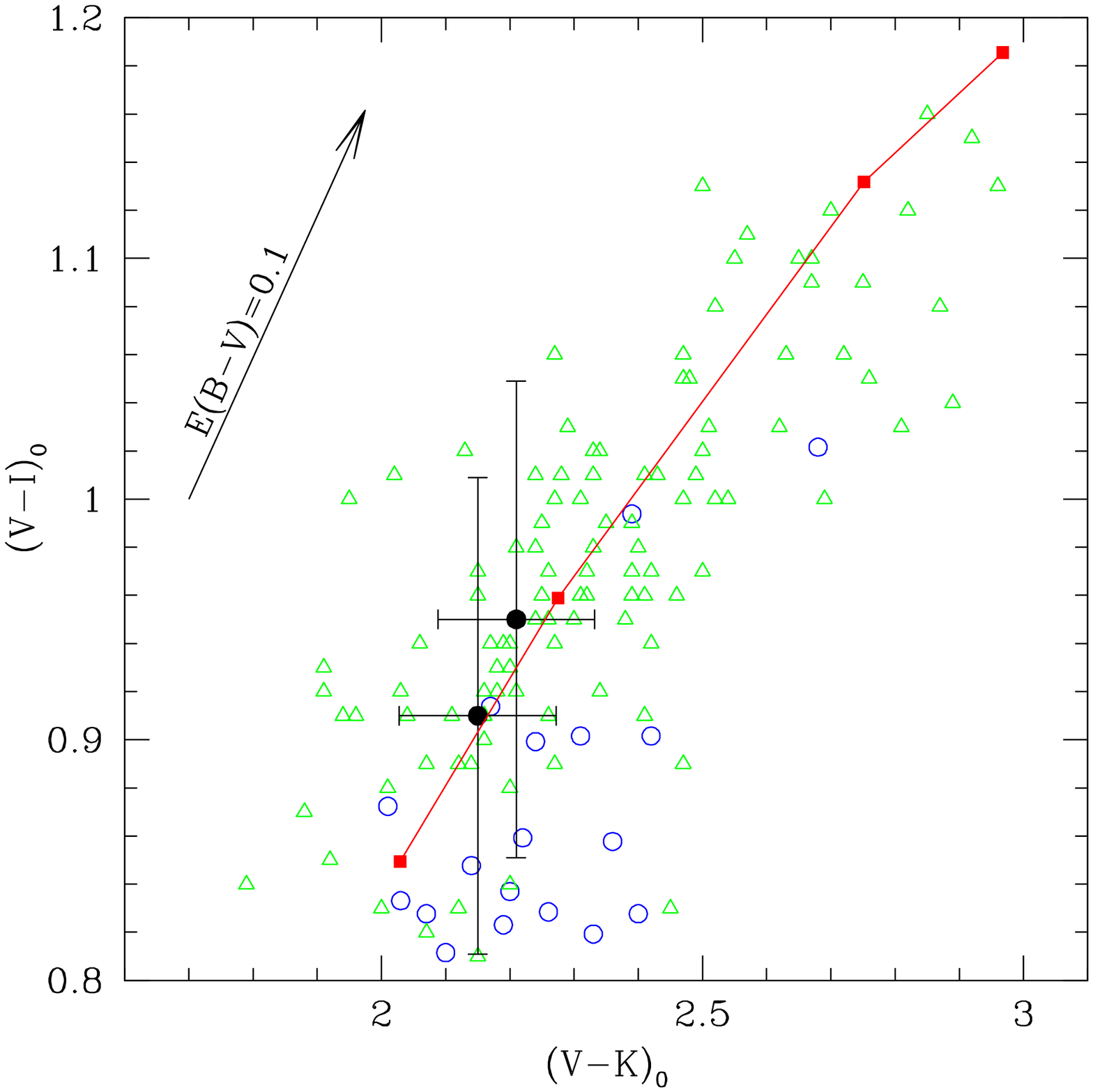}
}
\resizebox{\hsize}{!}{
\includegraphics[angle=0]{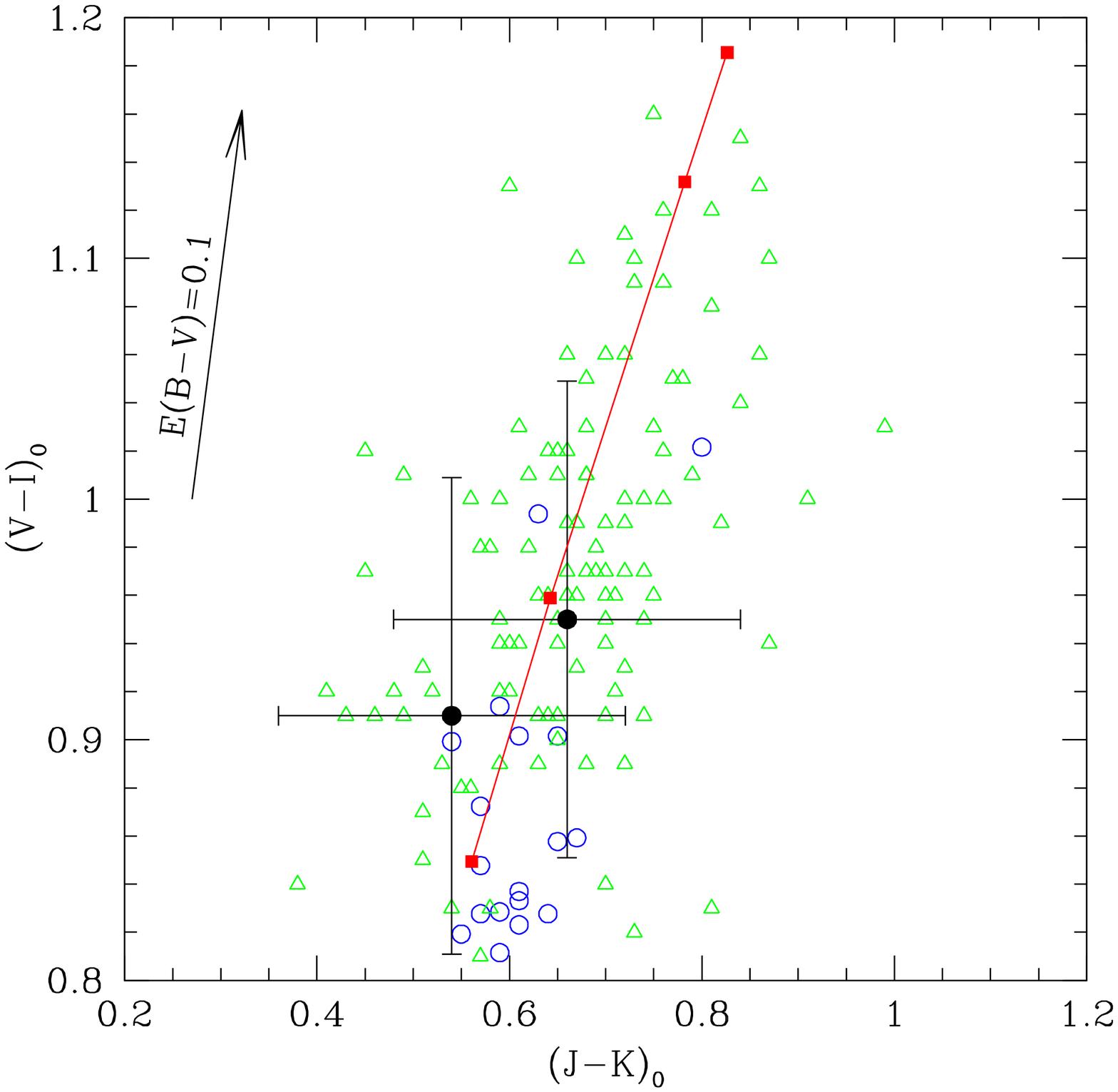}
}
\caption[]
	{Comparison of the optical and near-IR colours for the two globular
	cluster candidates in AM~1339-445 (filled dots with error-bars), 
	selected by \citet{sharina+05}, with globular 
	clusters in the Milky Way (open circles) and M31 (open triangles). We
	also plot a 12 Gyr constant age line for different metallicities 
        (filled
	squares): $\mathrm{[Fe/H]}=-2.25$, $-1.35$, $-0.33$ and $0.0$~dex, 
        going
	from blue to red colours \citep{maraston98,maraston05}. The reddening
	vector is plotted in the upper left for $\mathrm{E(B-V)}=0.1$~mag.}
   \label{fig:GCs}
\end{figure}

\citet{sharina+05} made a search for globular cluster candidates
in low surface brightness dwarf galaxies, based on WFPC2 HST archival images. 
In AM~1343-452 they found no globular cluster candidates, but AM~1339-445 
has two. Both of them are located on WF3 chip and are present in our 
ISAAC dataset. 

We have measured $J_s$ and $K_s$ magnitudes for these two globular cluster
candidates on our ISAAC images and report the reddening corrected
$J_{0}$ and $K_{0}$ magnitudes in Table~\ref{tab:GCs} 
together with optical magnitudes published by \citet{sharina+05}. 
We compare the colours of these candidates, plotted as filled symbols 
with error-bars, 
with those of the Milky Way (open circles; Aaronson, Malkan \& Cohen;
private communication from J.\ G.\ Cohen) and 
M31 \citep[open triangles;][]{barmby+00}
globular clusters in Fig.~\ref{fig:GCs}.  
For additional comparison, 
we also plot the iso-age line for a 12~Gyr 
single stellar populations (SSP) with metallicities ranging from 
$\mathrm{[Fe/H]}=-2.25$ to solar from \citet{maraston98,maraston05} models. 
The two globular cluster candidates have optical and near-IR colours similar 
to those of the old and metal-poor globular clusters in these two
large spiral galaxies. If we assume that their age is of the order of 12~Gyr,
then the AM~1339-445 clusters would have metallicities between 
$-1.8\la\mathrm{[Fe/H]}\la-1.4$.

%
%
\section{Discussion}
\label{sect:discuss}

Given the distance moduli and reddenings of the previous sections, we can
now use the $VI$ and $JK$ photometry of the upper-AGB candidates given 
in Table \ref{tab:AGBstars} to calculate M$_{bol}$ values for the stars.
For the $VI$ photometry we adopt the bolometric correction to the $I$
magnitude given by equation (2) of \citet{DA90}.  Although based on
globular cluster red giants, through the inclusion of red luminous giants
in 47~Tuc this equation is valid at least to $(V-I)_{0}$ $\approx$ 2.6.
It is thus applicable to all the stars in Table \ref{tab:AGBstars} with
the exception of the two reddest stars in AM~1343-452.  For these stars 
we have assumed that the relation can be extrapolated.  \citet{DA90} note
that there is no obvious metallicity dependance in the adopted 
equation, and that the dispersion about the fitted relation, 0.06 mag, is
consistent with the observational errors.
Using this relation, the three\footnote{In the following, wherever the sample
of upper-AGB stars is small, we give the luminosity of the three most 
luminous stars so as to indicate the likely location of the upper-AGB tip, 
and its uncertainty.} most luminous of the AM~1339-445 stars in 
Table \ref{tab:AGBstars} all have M$_{bol} \approx -4.4$, while for
AM~1343-452, the most luminous star has M$_{bol} \approx -4.75$ with
the two next most luminous stars both having M$_{bol} \approx -4.5$.

\begin{table*}
\caption{Absolute bolometric magnitudes for the brightest upper AGB stars in
AM~1339-445 and AM~1343-452 compared with those in the dE companions of the MW
and M31. For the MW dE companions we also list the time of the last 
significant star formation episode. The references to the literature data for
the AGB star magnitudes, and for the star formation histories 
(only for the MW dE companions) are given in
columns 3 and 5, respectively.}
\label{tab:AGB_in_dEs}
\centering
\begin{tabular}{lclcl}
\hline
\hline
galaxy & $\mathrm{M}_{bol}$ & $\mathrm{M}_{bol}$ reference& last SF & SFH reference\\
\hline
AM~1339-445  & $\approx -4.5$ & {\it this work} &  & \\
AM~1343-452  & $\approx -4.8$ & {\it this work} &  & \\
\hline
Leo~I     & $-5.1$ to $-4.2$ & \cite{menzies+02} & 1--2 Gyr & \cite{gallart+99} \\
Fornax    & $-5.2$, $-5.05$, $-4.8$ & \cite{demers+02} & 300--400 Myr & \cite{saviane+00} \\
Carina    & $-4.8$, $-4.7$, $-4.55$ & \cite{mould+82} & $\sim 3$~Gyr & \cite{hurley-keller+98} \\
Leo~II    & $-4.3$, $-4.1$, $-3.9$ &\cite{aaronson+mould85} & $\sim 7$~Gyr & \cite{mighell+rich96} \\
\hline
NGC~147  &  $-5.3$ & \cite{nowotny+03}  &   & \\
NGC~185  &  $-5.3$ & \cite{nowotny+03,kang+05} & & \\ 
And~II   &  $-4.7$, $-4.5$, $-4.25$ & \citet{kerschbaum+04} & & \\
\hline
\end{tabular}
\end{table*}

For the $JK$ photometry, there are a number of possible relations that 
provide bolometric corrections to the $K$ mag as a function of $(J-K)_{0}$
colour.  These include the 
relations given by \citet{BW84}, who list
relations based on oxygen-rich stars in the LMC and the Galaxy, and on
similar stars in 47~Tuc and the SMC\@.  Similarly, \citet{costa+frogel96} give
a relation derived from carbon stars in the LMC, while \citet{bergeat+02}
tabulate bolometric corrections as a function of $(J-K)_{0}$ 
based observations of Galactic carbon-rich giants.  We have investigated
these relations and find that, for the colour\footnote{We have adopted the 
weighted mean of the two $K$-band
observations in Table~\ref{tab:AGBstars} as the $K$ magnitude and used that
with the $J$ magnitude to form the $J-K$ colour.} range of the stars
in Table~\ref{tab:AGBstars}, the differences are
small ($\leq$0.10 mag) and are essentially independent of colour.  This is 
important because without further information we cannot be certain whether 
the candidate upper-AGB stars are M stars (oxygen-rich) or C stars 
(carbon-rich).  We note though that in metal-poor galaxies similar to those
studied here, stars with $(J-K)_{0}$ between $\sim$1.3 -- 1.5 and 2 are
typically found to be carbon-rich \citep[e.g.][]{cioni+habing03,raimondo+05}.  
In our upper-AGB candidate
lists there are 3 stars in each galaxy that fall in this colour range.  These
stars may well be C stars, although more precise photometry will be needed
to be certain of the classification.

For the bolometric corrections to the $JK$ photometry we adopt the 
relation of \citet{costa+frogel96} as it gives 
corrections that lie between those of \cite{BW84} and \cite{bergeat+02}.  
With this relation, the three most luminous stars (now based on the $JK$
photometry) in AM~1339-445 all have M$_{bol} \approx -4.6$, while for
AM~1343-452 the most luminous star has M$_{bol} \approx -4.85$ with
the next two most luminous having M$_{bol} \approx -4.7$ and $-4.5$.

As noted above, most upper-AGB stars are long period variables.  Consequently,
given the significant epoch difference between the $VI$ and $JK$ photometry,
we should not necessarily expect the most luminous stars in each set to
coincide.  Further, given the photometric errors and potential
systematic differences between the $VI$ and $JK$ bolometric corrections, we 
should also not necessarily expect the luminosities of the brightest star, 
or any combination of the luminosities of the brighter stars to agree.  
It is gratifying, therefore, to note that for each galaxy there is a 
reasonable degree of consistency between the two sets of data.  
For AM~1339-445 the brightest stars have M$_{bol} \approx -4.5 \pm 0.1$
regardless of whether the single most luminous, or the mean of the 3 most
luminous stars, are considered for both the $VI$ and $JK$ data sets.  One
star is also common to the most luminous three from the two sets.
Similarly, the most luminous AM~1343-452 star in our chosen sample 
has M$_{bol} \approx -4.8  \pm 0.1$ regardless of which data set
is used, and the mean of the three most luminous stars is $-4.65 \pm 0.1$
again independent of the photometry set.  As for AM~1339-445, there is
one star common to the most luminous three in the two data sets.

We have then our first clear result: {\it the upper-AGB reaches 0.2 -- 0.3
mag brighter in AM~1343-452 than it does in AM~1339-445}.  We note,
however, that the numbers of upper-AGB candidates are simply too small
to assert with any statistical significance (e.g.\ $\geq$2$\sigma$) that 
there are also relatively
more upper-AGB stars in AM~1343-452 than in AM~1339-445, despite the 12 vs 11
candidate ratio and a luminosity ratio for the areas surveyed of 
approximately 0.48 to 1.00.

How do these luminosities compare with those for upper-AGB stars in 
other dE galaxies?  We give in Table~\ref{tab:AGB_in_dEs} data for the
MW dE companions with clearly established upper-AGB populations. 
For each dE, we list the absolute bolometric magnitudes for the few 
brightest upper-AGB stars (full range for Leo~I) in column 2 with the 
reference to the literature 
source of the photometry in column 3. In most cases, the exception is
Leo~I where we have taken the values directly from \citet{menzies+02},  
the listed M$_{bol}$ 
values were obtained from the published photometry assuming distance moduli 
from \citet{mateo98} and the bolometric corrections of \citet{costa+frogel96}.
We also list in column 4 the time when the last
significant star formation episode ceased, based on literature data,
referenced in column 5.  These epochs are derived from 
analyses of star formation histories that are based on CMDs reaching 
well below the main sequence turnoff.

As for the dE companions to M31, the more luminous systems all contain notable 
populations of upper-AGB stars \citep[e.g.][and the references therein]{
battinelli+demers04a, battinelli+demers04b, demers+03}.  In particular, 
as indicated in Table~\ref{tab:AGB_in_dEs}, both 
NGC~147 and NGC~185 contain stars as bright as M$_{bol} \approx -5.3$
\citep{nowotny+03}; see also \citet{kang+05}.  However, unlike the majority 
of the MW's dE satellites,
the low luminosity dE satellites of M31 generally lack upper-AGB 
populations, with only And~II and And~VII (Cas) known to contain such 
stars \citep{aaronson+85,harbeck+04}.  \citet{kerschbaum+04} 
identify and give photometry for 7 carbon stars in And~II\@. Using the
$VI$ bolometric corrections of \citet{DA90} and distance and reddening
values from \citet{dacosta+00}, we give the bolometric magnitudes
for the three most luminous of these stars in Table~\ref{tab:AGB_in_dEs}.
These values are somewhat brighter than those discussed in Sect.\ 3.5.2 
of \citet{dacosta+00}, though it is unclear whether there are any stars 
in common.  No deep CMDs exist for the dE companions of M31, 
although, with Advanced Camera for Surveys (ACS) on board HST, it is now 
possible to obtain such data \citep{brown+03}.

We have then our second result: 
{\it the luminosities of the candidate upper-AGB 
stars found here for the Cen~A group dEs AM~1339-445 and AM~1343-452 are 
clearly quite comparable to those seen among Local Group dEs}.

\begin{figure}
\centering
\resizebox{\hsize}{!}{
\includegraphics[angle=0]{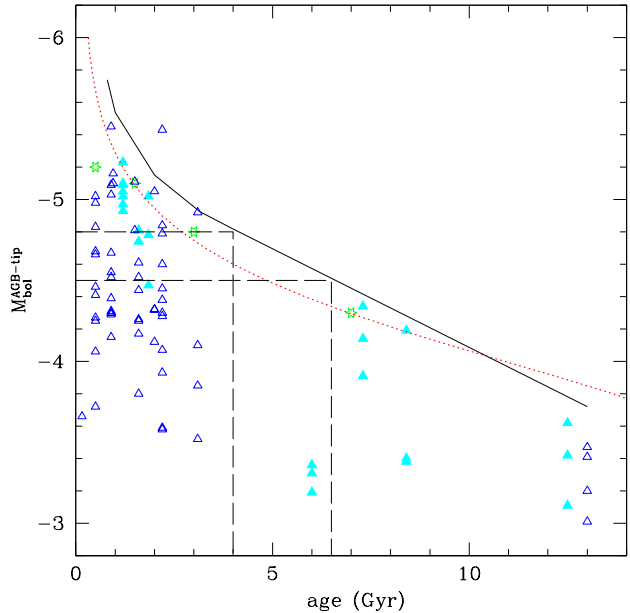}
}
\caption[]
	{ Our adopted relationship between AGB tip 
	luminosity and age is shown as a solid line. The LMC and SMC cluster
	data used to constrain this relation are shown as open (blue) triangles 
	and filled (cyan) triangles, respectively. The dotted (red) line 
	shows the theoretical AGB tip luminosity-age relationship from 
	\citet{stephens+frogel02}. 
	With open (green) star symbols we plot the MW dE galaxies from 
	Table~\ref{tab:AGB_in_dEs}, while dashed lines are used to 
	indicate the measured AGB tip bolometric magnitudes and the 
	inferred ages of the last significant star formation event 
	for the two dEs in Cen A group.}
   \label{fig:ageAGBtip}
\end{figure}

We turn now to estimating the age of the most luminous of the upper-AGB stars
in each dE\@.  The relatively small numbers of stars in our
candidate lists, combined with the rapidity of this evolutionary phase, means
that the age estimates derived below should strictly be considered upper 
limits: it is possible that a larger sample (assuming it was possible to 
generate one) might reveal more luminous stars.  For this same reason, the 
derived ages should be considered as limits on the epoch of the most recent 
episode of significant star formation in these galaxies: star formation 
may well have continued past these eras at a sufficiently low rate that the 
upper-AGB is simply not populated.  A further complication is that modeling
the thermally-pulsing phase of AGB evolution is difficult to carry out in 
detail, and the results are sensitive to the mass-loss prescriptions and
the details of the mixing processes that occur during the thermal pulses,
as well as to changes in molecular opacities during the third dredge-up 
\citep{marigo02}.  
Thus any theoretical calibration of AGB tip luminosity with age
and abundance is necessarily uncertain.
The star clusters of the Magellanic Clouds, however, provide
a set of objects with which an empirical calibration of the bolometric
magnitude of the AGB tip as a function of age can be generated.  While these
clusters, particularly those in the LMC, are somewhat more metal-rich than 
the dE stars considered here, the theoretical models suggest that the 
dependance on metallicity at fixed age is not large.  For example, using
the isochrones of \citet{girardi+02}, the difference in AGB tip luminosity
at 2 Gyr for metallicities Z=0.0004 and Z=0.004 is only $\sim$0.3 mag (the 
more metal-poor brighter), with the difference decreasing with increasing age.
Conversely, for the same isochrones, the difference in the AGB tip
luminosity between 1 and 3 Gyr is $\sim$0.6 mag for both metallicities. 

Using the bolometric magnitudes tabulated by \citet[][ see also \cite{fer+95,
fer+04}]{frogel+90}, LMC and SMC distance moduli of 18.5 and 18.9, 
respectively, and cluster ages from a variety of compilations 
\citep[e.g.][]{dacosta02}, we have constructed  an empirical relationship
between AGB tip luminosity and age, for ages older than approximately
1 Gyr.  This relation is primarily based on those clusters with extensive
near-IR observations, e.g.\ NGC~1783 \citep{mould+89}, and naturally, given 
the lack of such clusters in the LMC, it is dependant solely on SMC 
clusters for ages beyond 3 Gyr. For ages less than this, however, there 
is no obvious difference between the two sets of clusters. Further, the 
relation gives results consistent with the (AGB tip luminosity, age of 
last significant episode of star formation) results quoted above for the 
MW dE companions. 
Our adopted relation is shown as the solid line in 
Fig.~\ref{fig:ageAGBtip}, where each vertical sequence of points 
at fixed age represents all the upper AGB stars for a given cluster. 
The MW dE companion data from Table~\ref{tab:AGB_in_dEs} 
are plotted with open star symbols. For comparison, with 
dotted line we plot the relation between the AGB tip
luminosity and age from \citet{stephens+frogel02}, determined by 
using the stellar evolution models of \citet{bertelli+94} and the mass-loss 
prescription of \citet{vw93}. 

Using our empirically derived relation, and the assumed AGB 
tip luminosities 
($-4.5$ for AM~1339-445, and $-4.8$ for AM~1343-452) we derive epochs of the
last significant episode of star formation in these Cen~A group dEs as
$6.5 \pm 1$ and $ 4\pm 1$ Gyr, respectively (dashed lines in
Fig.~\ref{fig:ageAGBtip}). The uncertainty in age is determined by 
$\pm 0.1$~mag uncertainty in AGB tip bolometric magnitude, but we note 
that it does not include uncertainty in the relation between the AGB tip 
luminosity and age. Clearly use of the  \citet{stephens+frogel02} relation would
result in somewhat younger ages for the last episode of significant star
formation. Nevertheless, we conclude that AM~1339-445
is comparable to the MW satellite Leo~II, while AM~1343-452 is 
comparable to Carina. 

While the ubiquitous occurrence of RR~Lyrae variables shows that the 
outlying dE satellites of the MW contain at least some old (age $\geq$ 10 Gyr)
stars \citep[e.g.][]{held+01}, the stellar populations of these systems,
particularly Fornax and Leo~I, are dominated by stars of intermediate-age.
It is therefore of interest to attempt to estimate the relative importance
of the intermediate-age populations in the Cen~A group dEs.  Clearly
the lack of observations reaching substantially fainter than the tip of
the red giant branch means that we cannot provide detailed estimates
of the star formation histories \citep[cf.][]{gallart+99}.  Nevertheless,
we can use simple models to provide some information on the relative 
importance of the intermediate-age stars.  Specifically, we assume that
the dwarfs are made up of two discrete populations: one which is old,
assumed to be 13 Gyr in age, and one which is of intermediate age with
the same metallicity.  The intermediate-age population fraction is then
estimated from the numbers of candidate upper-AGB stars (11 for AM~1339-445,
12 for AM~1343-452), which are presumed to come only from the 
intermediate-age population, as a proportion of the number of red giants
between the tip of the RGB and 0.3 mag fainter (in $I$).  These stars
come from both populations.  We denote this ratio by $P_{IA}$.
Using the CMDs of Fig.\ \ref{fig:RGBsVI} we find $P_{IA}$ $\approx$ 
0.05-0.10 for both AM~1339-445 and AM~1343-452.

We use models kindly provided by Claudia Maraston to interpret these
ratios.  The models (cf.\ \citet{maraston98,maraston05}; see also
\citet{fer+04}) estimate the energetics of any post-main-sequence phase
by using the so-called fuel consumption theorem \citep{renzini+buzzoni86}, 
and have
been calibrated with the integrated colours of Magellanic Cloud clusters.
Using the models for $\mathrm{[Z/H]} = -1.35$, 
which is appropriate for the dEs
given the mean metallicities derived in Sect.\ \ref{sect:results_optical},
we calculate values of $P_{IA}$ for ages of the intermediate-age
component of 1, 2, 4, 6 and 9 Gyr, and for intermediate-age population
fractions varying from 0.0 (i.e.\ old stars only) to 1.0 
(i.e.\ intermediate-age stars only).  In all cases we include in the
calculation of the relative number of red giants between $I$(TRGB) and
$I$(TRGB+0.3) not only first ascent red giants, but also early-AGB stars
in the same $I$ magnitude range.

The models indicate that if we assume an age of $\sim$6.5 Gyr for the
intermediate-age population in AM~1339-445, and $\sim$4 Gyr for AM~1343-452,
then the observed values of $P_{IA}$ for the dEs imply
intermediate-age population fractions of $\sim$15\% for both galaxies.
It is not easy to give an uncertainty for this value, but it most likely
a lower limit on the true fraction.  For example,
the number of upper-AGB candidates is likely to be a lower limit on
the actual number, given the selection process and the photometric errors.
Similarly, since the models show that relative number of upper-AGB stars
per RGB star (first ascent and early-AGB) decreases with increasing age, 
including an additional intermediate-age population older than that used
would result in a larger overall intermediate-age population fraction.
It is unlikely, however, that the true intermediate-age population 
fractions for these dEs are more than a factor of two higher than the 
limit given.  

At $\sim$15\%, or even at $\sim$30\%, these intermediate-age
population fractions are notably lower than for the outlying MW dE satellites
Fornax and Leo~I, where the intermediate-age population fractions are
dominant.  Even for Leo~II, the results of \citet{mighell+rich96}
suggest an intermediate-age population fraction of $\sim$40-50\%.  The
Cen~A group dEs appear instead to be more similar to the outlying 
lower luminosity dE companions of M31.  

As regards the comparison of these Cen~A group dEs with dEs in the Local
Group, two further points can be made.  First, we have already noted that 
the $(B-R)_0$ colours of these two galaxies
are typical of those for dEs \citep{JBF00}.  Second, using the photometry of 
\citet{JBF00} and our moduli, we estimate that the integrated absolute $V$ 
magnitudes are $-12.6$ and $-11.5$, for AM~1339-445 and 
AM~1343-452, respectively.  With these magnitudes and the average 
metallicities of $-1.4$ and $-1.6$ derived above, both
galaxies follow closely the luminosity-metallicity relation defined 
by the Local Group dEs \citep{caldwell+98}.

%
%

\section{Summary and Conclusions}

We have presented an analysis of the red giant populations of two
dE galaxies in the Cen~A group, AM~1339-445 and AM~1343-452, 
using a combination of near-IR and
optical data.  Both dEs are distant companions of Cen~A, the dominant
galaxy of the group. 

Using the luminosity of the tip of the RGB we have measured distance moduli
of both galaxies, 
$(\mathrm{m}-\mathrm{M})_0(\mathrm{AM1339-445})=27.74 \pm 0.20$, 
and $(\mathrm{m}-\mathrm{M})_0(\mathrm{AM1343-452})=27.86 \pm 0.20$, 
which are in good agreement with previously published values 
\citep{k+02,JFB00}. The mean colour of the upper RGB stars is used 
to determine the mean metallicities of
$\langle \mathrm{[Fe/H]}\rangle =-1.4 \pm 0.2$ for AM~1339-445 and 
$\langle \mathrm{[Fe/H]}\rangle =-1.6 \pm 0.2$ for AM~1343-452.  
The integrated colours of these two dEs are similar to those of other 
dEs and they follow the same
luminosity-metallicity relation of the LG dEs.

We find evidence for the presence of intermediate-age
upper-AGB stars in both galaxies, with the most luminous of these stars 
being 0.2-0.3 mag brighter in AM~1343-452, than in AM~1339-445.  The 
luminosities of these stars
indicate that significant star formation continued in AM~1343-452 until
an age of $\sim$4 Gyr as against $\sim$6.5 Gyr in AM~1339-445.  In this
respect these Cen~A group dEs are similar to the outlying dE satellites
of the Milky Way.  However, we estimate that the fraction of the total 
population that is of intermediate-age is perhaps $\sim$15\%, which
is significantly less than the dominant intermediate-age populations found in 
outlying Milky Way dE satellites such as Fornax and Leo~I. 

With only two galaxies it is premature to draw any definite conclusions
regarding our long term goal of investigating the role of environment
on the evolution of dE galaxies in the Cen~A group 
(cf.~Sect.~\ref{sect:Intro}).
Nevertheless, it is interesting that despite the rather large distance of
both dEs from Cen~A, their intermediate-age populations are small and
relatively old, particularly when compared to the outer dE satellites of 
the Milky Way.  We must, however, await similar analyses for additional dEs 
in this group before drawing any inferences from this result. 

%
%

\begin{acknowledgements}
We are grateful to our referee, Ivo Saviane, for many useful suggestions 
which improved the presentation.
We thank service mode support at Paranal for conducting the observations
and Dr.\ Claudia Maraston for providing details from her models.  
GDaC also thanks Prof.\ Roger 
Davies, head of Astrophysics at Oxford University, for hosting a sabbatical 
visit during which this paper was completed.  The research has been supported
in part by funds from the Australian Research Council through Discovery
Project grant DP0343156. BB thanks the Swiss National
Science Foundation for financial support.
This publication makes use of data products from the 
Two Micron All Sky Survey, which is a joint project of the University of 
Massachusetts and the Infrared Processing and Analysis Center/California 
Institute of Technology, funded by the National Aeronautics and Space 
Administration and the National Science Foundation. This research has 
made use of NASA's Astrophysics Data System Bibliographic Services.
\end{acknowledgements}

%
%

\bibliographystyle{aa}
\bibliography{MS4249.bbl}

%
%

\end{document}